\documentstyle[12pt,twoside]{article}

\newcommand{\D}{\displaystyle}
\newcommand{\T}{\textstyle}

\newcommand{\unit}[1]{\mbox{\ #1}}
\newcommand{\subs}[1]{{\mbox{\scriptsize\it #1}}}
\newcommand{\comment}[1]{}

\newcommand{\str}{\rule{0ex}{2.7ex}}
\newcommand{\go}{\, {}_\sim^{>} \,}
\newcommand{\lo}{\, {}_\sim^{<} \,}

\newcommand{\be}{\begin{equation}}
\newcommand{\ee}{\end{equation}}
\newcommand{\acal}{{\cal A}}

\newcommand{\mcal}{{\cal M}}

\newcommand{\tcal}{{\cal T}}

\newcommand{\gm}{\gamma_{-}}
\newcommand{\gp}{\gamma_{+}}
\newcommand{\Res}{\mbox{BW}}
\newcommand{\strich}[1]{#1  \! \! \slash}
\newcommand{\BW}{{\mbox{BW}}}
\newcommand{\br}{\mbox{BR}}

\newcommand{\Li}{\mbox{Li}_{2}}

\newcommand{\tr}{\mbox{tr}}
\newcommand{\kst}{{K^{\star}}}
\renewcommand{\Re}{\mbox{Re}}

\newcommand{\eqn}[1]{Eqn.~(\ref{#1})}
\newcommand{\fig}[1]{Fig.~\ref{#1}}
\newcommand{\sect}[1]{Sec.~\ref{#1}}
\newcommand{\app}[1]{App.~\ref{#1}}
\newcommand{\tab}[1]{Tab.~\ref{#1}}

\begin{document}
\everymath={\displaystyle}
\thispagestyle{empty}
\vspace*{-2mm}
\thispagestyle{empty}
\noindent
\hfill TTP94--5\\
\mbox{}
\hfill  June 1994  \\
\vspace{1cm}
\begin{center}
  \begin{Large}
  \begin{bf}
SHORT AND LONG DISTANCE EFFECTS\\
IN THE DECAY $\tau \to \pi \nu_\tau (\gamma)$
   \\
  \end{bf}
  \end{Large}
  \vspace{0.8cm}
  \begin{large}
   Roger Decker and Markus Finkemeier\\[5mm]
    Institut f\"ur Theoretische Teilchenphysik\\
    Universit\"at Karlsruhe\\[2mm]
    76128 Karlsruhe\\ Germany\\
  \end{large}
\bf{1cm}
  {\bf Abstract}
\end{center}
\begin{quotation}
\noindent
We calculate the radiative corrections to the decays
$\tau\to M \nu_\tau$ and $\pi \to l \nu_\l$,
where the meson $M$ is  $M=\pi$ or $K$ and the lepton $l$ is $l = e$
or $\mu$.
We perform a complete calculation, which includes internal
bremsstrahlung and structure dependent radiation in the radiative
decays and point meson, hadronic structure dependent and short
distance contributions in the virtual corrections.
Our result for the radiative correction to the ratio
$\Gamma(\tau\to\pi\nu_\tau(\gamma))/ \Gamma(\pi\to\mu\nu_\mu(\gamma))$
is
$\delta R_{\tau/\pi} = \left(0.16_{\T - 0.14}^{\T+
0.09}\right)  \%$.
For the ratio
$\Gamma(\tau\to K\nu_\tau(\gamma))/ \Gamma(K\to\mu\nu_\mu(\gamma))$,
we obtain
$\delta R_{\tau/K } = \left(0.90_{\T - 0.26}^{\T + 0.17}\right)
\%$.
For completeness
we have also calculated the ratio of the electronic and muonic decay
modes of the pion.
\end{quotation}

%
%
\section{Introduction}
In the calculation of radiative corrections to semileptonic
(semihadronic) decays such as $\tau\to\pi\nu_\tau$ and
$\pi\to\mu\nu_\mu$, one faces three different problems
\cite{liste1,Kin59}.
As usual in radiative corrections, there are divergences,
viz. first the infra-red (IR) divergences and second the
ultra-violet (UV) divergences. The third problem, however, which is
the central issue of this paper, is the treatment of the strong
interaction.

The IR divergences are removed as usual by considering either
radiative decays with hard photons (eg.\ , $\Gamma(\tau\to\pi\nu_\tau
\gamma$) with $E_\gamma > E_0$)
or inclusive rates for decays into final states with and without photon
(eg.\ $\Gamma(\tau\to\pi\nu_\tau) +
\Gamma(\tau\to\pi\nu_\tau\gamma)$).

The UV divergences are removed by renormalization. The decay rate of
$\pi\to\mu\nu_\mu$ is proportional to the pion decay constant $f_\pi$.
In principle $f_\pi$ is determined by the parameters of the standard
model, but since we are not able to solve the nonperturbative regime
of
QCD, $f_\pi$ has to be considered as an additional free parameter
which has to be extracted from experimental data and therefore has to be
renormalized order by order in perturbation theory.
On the other hand, ratios as
\be
   R_{\tau/\pi} := \frac{\Gamma(\tau\to\pi\nu_\tau(\gamma))}
   {\Gamma(\pi\to\mu\nu_\mu(\gamma))}
\ee
and
\be
   R_{e/\mu} := \frac{\Gamma(\pi\to
   e\nu_e(\gamma))}{ \Gamma(\pi\to\mu\nu_\mu(\gamma))}
\ee
are independent of $f_\pi$ and therefore
can be predicted by theory. Technically the UV divergences
cancel in these ratios.

For a systematic treatment of the issue of strong interaction, a
separation into different energy regimes should be made.
If all momenta are very small compared to a typical hadronic
resonance scale such as $m_\rho$, the matrix elements are fixed by
low energy theorems of QCD.
In this low energy regime the pion behaves like a
pointlike particle, and its interactions with the photon are
determined by scalar QED. The high energy regime, on the other hand,
is dominated by the short distance corrections to the weak interaction,
ie.\ by photonic corrections acting at the quark level. These two
regimes are not adjacent to each other, but rather there is an
intermediate region which is dominated by non-perturbative strong
interaction, viz. by the physics of hadron resonances such as the
$\rho$ and the $a_1$ particles.

In a previous paper \cite{Dec93b}, we have calculated the corrections
to $R_{\tau/\pi}$ within a model with an effective point pion field.
Defining the radiative correction $\delta R_{\tau/\pi}$ by
\be \
   R_{\tau/\pi}
   = R_{\tau/\pi}^{(0)} \Big(1 + \delta R_{\tau/\pi} \Big)
\ee
where
\be
   R_{\tau/\pi}^{(0)}
   = \frac{1}{2} \frac{m_\tau^3}{m_\pi m_\mu^2}
     \frac{\left(1- m_\pi^2 /  m_\tau^2 \right)^2}
     {\left(1- m_\mu^2 /  m_\pi^2 \right)^2}
\ee
denotes the prediction to the order $O(\alpha^0)$,
we found a radiative correction
\be \label{eqnpm}
   \delta R_{\tau/\pi}^{(P.M.)} = +1.05\%.
\ee
(P.\ M.\  denotes point meson).
We added to this result another $+0.17 \%$ arising from
hadronic structure dependent radiation.
So what is missing in \cite{Dec93b} are the hadronic structure
dependent
effects in the virtual corrections and the short distance corrections.

As for the short distance correction, it was shown in
Ref.~\cite{Sir82} that the leading $O(\alpha)$ correction in the limit
of a large $Z$ boson mass, $m_Z^2\to\infty$, is
\be
    {\mcal_0} \to \left[ 1 + \frac{\alpha}{\pi}\ln \frac{m_Z}{\mu}
     \right] \mcal_0
\ee
where $\mcal_0$ denotes the Born amplitude, $m_Z$ acts as a cut-off
and $\mu$
is an unspecified mass scale characteristic of the process.
The difficulty is of course to find identify the scale
$\mu$. If the relevant scale is $\mu_1$ for the decay
$\tau\to\pi\nu_\tau (\gamma)$ and $\mu_2$ for
$\pi\to\mu\nu_\mu(\gamma)$, the
short distance contribution to the radiative correction of
$R_{\tau/\pi}$ is
\be
   \frac{2 \alpha}{\pi} \left(\ln\frac{m_Z}{\mu_1} -
   \ln\frac{m_Z}{\mu_2}\right)
   = \frac{2 \alpha}{\pi} \ln \frac{\mu_2}{\mu_1}
\ee
and so only the difference in the scales $\mu_2$ and $\mu_1$ is
relevant for the correction to $R_{\tau/\pi}$.
In Ref.~\cite{Mar88} Marciano and Sirlin give an estimate for $\delta
R_{\tau/\pi}$ which is based on the short distance contribution only,
\be
   R_{\tau/\pi}^{(s.d.)}  =   R_{\tau/\pi}^{(0)}
    \frac{1 + 2 \alpha/\pi \ln(m_Z/m_\tau)}
   {1 + \frac{3}{2}(\alpha/\pi) \ln(m_Z/m_\pi)
      + \frac{1}{2}(\alpha/\pi) \ln (m_Z/m_\rho)}
\ee
(s.\ d.\ denotes short distance)
or
\be \label{eqnms1}
  \delta R_{\tau/\pi}^{(s.d.)} =
  \frac{2\alpha}{ \pi} \ln
    \left(
    \frac{m_\pi^{3/4} m_\rho^{1/4}}{m_\tau}
    \right)
   = -0.98 \%
\ee
This result differs from the effective point meson result.
But note that in the case of the pion decay this estimate extends the
short distance physics to a very small scale of $\mu_2 =
m_\pi^{3/4} m_\rho^{1/4}= 214 \unit{MeV}$.

In a subsequent paper \cite{Mar92}, Marciano included both the point
meson and the short distance corrections to $\pi\to l \nu_l (\gamma)$,
matched at the scale $m_\rho$. For the $\tau$ decay
$\tau\to\pi\nu_\tau(\gamma)$, still only the short distance
corrections were included.
Reexpressing his prediction in terms of $\delta R_{\tau/\pi}$, we
obtain from \cite{Mar92,Mar91}
\be \label{eqnmarciano}
   \delta R_{\tau/\pi} = (0.55  \pm 1.) \%
\ee
where the $\pm 1 \%$ is the author's estimate of the missing long
distance corrections of $O(\alpha)$ to $\tau\to\pi\nu_\tau(\gamma)$.
In a recent papper \cite{Mar93}, the authors have further improved the
calculation of the radiative pion decay by including the leading
hadronic structure dependent effects (both in the radiative decay and
in the virtual corrections) and by including the leading two-loop
effects. Using the pion decay constant $f_\pi$ extracted from this
calculation to predict the tau
decay, they obtain a new prediction for the rate, which can be
rewritten in terms of $R_{\tau/\pi}$ as
\be \label{eqnms2}
    \delta R_{\tau/\pi} = (0.67  \pm 1.) \%
\ee
where again the $\pm 1\%$ estimates the long distance
corrections to $\tau\to\pi\nu_\tau(\gamma)$, which are still missing.

Comparing the numbers in (\ref{eqnpm},\ref{eqnms1},\ref{eqnmarciano},%
\ref{eqnms2}), it becomes clear that a complete and systematic
calculation of the full $O(\alpha)$ corrections would be important
in order to obtain a reliable prediction.
This is what we will present in this paper.

We have performed a systematic calculation of the
radiative corrections to $R_{\tau/\pi}$ which includes all relevant
contributions. In the calculation of the loops, we separate the
integration over the momentum of the virtual photon $k$ into the long
distance region with $0 \leq |k^2| \,{}_\sim^{<}\, \mu_{cut}^2$ and
into the short distance region with $\mu_{cut}^2 \,{}_\sim^{<}  |k^2|
\, \leq m_Z^2$.
The matching scale $\mu_{cut}$, which separates long and
short wavelengths, should be of the order $\mu_{cut} \sim O(1
\unit{GeV})$. The long distance part includes the effective point
meson and the hadronic structure dependent corrections. The latter is
obtained by modifying the scalar QED coupling of the photon to the pion
by vector meson dominance, and by adding loops which are proportional
to the form factors $F_V$ and $F_A$ which determine the hadronic
structure dependent radiation.

For the short distance
corrections we first consider the leading
logarithm,
and second we give a
complete calculation based on the parton model.

Our paper is organized as follows: In Sec.~2 we briefly review the
amplitudes for the radiative decays and the parametrizations of the two
form factors appearing in the hadronic structure dependent radiation.
In Sec.~3 we make some general remarks on our treatment of the virtual
corrections, concerning the $W$ boson propagator and the separation
into long and short distances. In Sec.~4 we calculate the point meson
contribution. We have published the results for the radiative
corrections  in a model with a pointlike pion in \cite{Dec93b}.
Nevertheless we
repeat them here for completeness and in order to give some more
details and
intermediate results, which
are needed later in the calculation in order to combine the point meson
correction with the other contributions.
In Sec.~5 we consider the leading
logarithm of the short distance contribution, and in Sec.~6 we
calculate the non-leading corrections. In Sec.~7 we calculate the
hadronic structure dependent loops . Then in Sec.~8 we explain how the
different contributions combine to give the final result, which we
evaluate numerically in Sec.~9. A summary and concluding remarks are
given in Sec.~10.
%
\section{The Radiative Decays}
\label{secrad}
%
We consider the decays $\tau\to M
\nu_\tau (\gamma)$ and $M \to l \nu_l (\gamma)$, where the meson $M$
is the pion $M = \pi$ or the kaon $M = K$, and the lepton $l$ is the
muon or the electron, $l = \mu$ or $e$.
The Born
amplitudes are given by
\begin{eqnarray}
   \mcal_0(\tau(s)\to M(p) \nu_\tau(q))  & = &
    - G_F V_M f_M [\bar{u}_\nu(q) \strich{p} \gm u_\tau(s)]
\nonumber \\
\nonumber \\
   \mcal_0(M(p) \to l(s) \nu_l(q))  & = &
    G_F V_M f_M [\bar{u}_\nu(q) \strich{p} \gm v_l(s)]
\end{eqnarray}
where
\be
   V_\pi = \cos \theta_C \qquad \qquad V_K = \sin \theta_C
\ee

The matrix elements of the radiative decays $\tau\to M
\nu_\tau\gamma$ \cite{liste2,Dec93}
and
$M\to l \nu_l \gamma$ \cite{liste3} can be written as the sums of
two contributions,
the internal bremsstrahlung (IB) and the structure dependent radiation
(SD)
\be
   \mcal [\tau^-(s) \longrightarrow \nu_\tau(q) M^-(p) \gamma(k)]
   = \mcal_{IB} + \mcal_{SD}
\ee
where
\begin{eqnarray}
   \mcal_{IB} & = & - G_F V_M e f_M m_\tau \left[ \bar{u}_\nu(q) \gp \left(
      \frac{p \cdot \epsilon}{p \cdot k} + \frac{\strich{k}
      \strich{\epsilon}}{2 s \cdot k} - \frac{s \cdot \epsilon}{s \cdot k}
      \right) u_\tau(s) \right] \nonumber \\
   \mcal_{SD} & = & - \frac{G_F V_M e }{\sqrt{2}} \Bigg\{ i \epsilon_{\mu \nu
\rho
      \sigma}
    \Bigg[ \bar{u}_\nu(q) \gamma^\mu \gm u_\tau(s) \Bigg]
   \epsilon^\nu k^\rho p^\sigma \frac{F^{(M)}_V((k+p)^2)}{m_M}
\nonumber \\
& & \qquad
+ \Bigg[ \bar{u}_\nu(q) \gp \Big( (p \cdot k) \strich{\epsilon} -
      (\epsilon \cdot p) \strich{k} \Big) u_\tau(s) \Bigg]
      \frac{F^{(M)}_A((k+p)^2)}{m_M}
      \Bigg\}
\end{eqnarray}
Similarly
\be
   \mcal [M^+(p) \longrightarrow l^+(s) \nu_l(q) \gamma(k)]
   = \mcal'_{IB} + \mcal'_{SD}
\ee
where
\begin{eqnarray}
   \mcal'_{IB} & = & - G_F V_M e f_M m_l \left[ \bar{u}_\nu(q) \gp \left(
      \frac{p \cdot \epsilon}{p \cdot k} - \frac{\strich{k}
      \strich{\epsilon}}{2 s \cdot k} - \frac{s \cdot \epsilon}{s \cdot k}
      \right) v_l(s) \right] \nonumber \\
   \mcal'_{SD} & = &  \frac{G_F V_M e }{\sqrt{2}} \Bigg\{ i \epsilon_{\mu \nu
\rho
      \sigma}
    \Bigg[ \bar{u}_\nu(q) \gamma^\mu \gm v_l(s) \Bigg]
   \epsilon^\nu k^\rho p^\sigma \frac{F^{(M)}_V((k-p)^2)}{m_M}
\nonumber \\
& & \qquad
+ \Bigg[ \bar{u}_\nu(q) \gp \Big( (p \cdot k) \strich{\epsilon} -
      (\epsilon \cdot p) \strich{k} \Big) v_l(s) \Bigg]
      \frac{F^{(M)}_A((k-p)^2)}{m_M}
      \Bigg\}
\end{eqnarray}

The IB part is fixed by the QCD low energy theorems. It
is exactly identical to the one which would be obtained
for an elementary pointlike pion field, with electromagnetic
interactions determined by scalar QED. The IB amplitude does not
contain any unknown parameters beyond the meson decay constant $f_M$.
The SD contribution, on the other hand, involves the two form factors
$F^{(M)}_V(t)$ and $F^{(M)}_A(t)$, which describe the effects of
non-perturbative
strong interactions. Crossing symmetry implies that both the meson and
the tau decays are described by the same analytical functions of the
momentum transfer $t$, the difference being that in the $\tau$
decay, $t$ varies from $m_M^2$ up to $m_\tau^2$, and in radiative meson
decay, $t = 0 \dots m_M^2$.
In Ref.~\cite{Dec93} we have parametrized these form factors.

For the pionic case $M=\pi$, we gave
\be
    F^{(\pi)}_A(t) = F^{(\pi)}_A(0) \Res_{a_1}(t)
    \label{eqnfa}
\ee
and
\be
   F^{(\pi)}_V(t) = F^{(\pi)}_V(0)  \left[\Res_\rho(t) +
        \sigma \Res_{\rho'}(t) + \rho \Res_{\rho''}(t) \right]
        \frac{1}{1 + \sigma + \rho}
\ee
$F^{(\pi)}_A(0)$ has been measured in radiative pion decay
\cite{RPP92},
\be \label{eqn266}
   F^{(\pi)}_A(0)
   = +0.0116 \pm 0.0016.
\ee
whereas $F^{(\pi)}_V(0)$ is related to the axial anomaly and predicted
to be
\be
   F^{(\pi)}_V(0) =  \frac{m_\pi}{4 \sqrt{2} \pi^2 f_\pi} = + 0.0270
\ee
Note that here we use a relative sign $s := \mbox{sign}(f_\pi F_V(0))$
which is positive in our conventions, $s = + 1$. In fact in
\cite{Dec93}, we used $s = -1$.
However, as we have explained in \cite{Dec93A}, we now believe that
$s=+1$ is the physical choice, which we will therefore use throughout
this paper.

The Breit-Wigner resonance factors $\Res_X(t)$ are normalized
according to $\Res_X(0) = 1$ and involve energy dependent widths
\be
   \Res_X(t) = \frac{m_X^2}%
      {m_X^2 - t - i m_X \Gamma_X(t)}
\ee
or constants widths
\be
  \label{eqnf2}
   \Res_X(t) = \frac{m_x^2 - i m_X \Gamma_X}{m_X^2 - t - i m_X \Gamma_X}
\ee
For the $a_1$ resonance,
an energy dependent width based on the three
pion and the pion-rho phase space has been calculated
in \cite{Kue90},
\be
   \Gamma_{a_1}(t) = \frac{g(t)}{g(m_{a_1}^2)} \Gamma_{a_1}
\ee
with
\be
   g(t) = \left\{ \begin{array}{lll}
      0 & \mbox{\ \ if\ \ } t < 9 m_\pi^2 \\
      4.1 (t - 9 m_\pi^2)^3 [1 - 3.3 (t - 9 m_\pi^2) + 5.8
      (t - 9 m_\pi^2)^2] & \mbox{\ \ if\ \ } 9 m_\pi^2 \leq t < (m_\rho
      + m_\pi)^2 \\
      t(1.623 + 10.38/t^2 - 9.32 / t^4 + 0.65/t^6) &
      \mbox{\ \ else\ \ }
   \end{array} \right. 
\ee
(all numbers in appropriate powers of $\unit{GeV}$)

For the $\rho$, $\rho'$ and $\rho''$ resonances, an energy dependent width
\be
   \Gamma_X(t) = \left\{ \begin{array}{ll}
       0 & \mbox{\ \ if\ \ } t \leq 4 m_\pi^2 \\
       \D \frac{m_X}{\sqrt{t}} \left( \frac{\sqrt{t - 4 m_\pi^2}}%
       {\sqrt{m_X^2 - 4 m_\pi^2}} \right)^3 \Gamma_X &
       \mbox{\ \ else\ \ }
       \end{array} \right.
\ee
can be derived from the P-wave two body phase space.

We fixed
the two parameters $\sigma$ and $\rho$ in \cite{Dec93} by
employing four
constraints, viz. the QCD theorem on $\lim_{t\to\infty}
F_{\pi\gamma}(t)$ \cite{Lep79}
(where $F^{(\pi)}_V(t) = - m_\pi F_{\pi\gamma}(t) /
\sqrt{2}$ \cite{Vak58}), the measurement of the slope
$F'_{\pi\gamma}(t=0)$ and the
widths $\Gamma_{\rho\to\pi\gamma}$ and $\Gamma_{\rho'\to\pi\omega}$.
The result was
\begin{eqnarray}
   \sigma & = & + 0.136 \nonumber \\
   \rho & = & - 0.051
\end{eqnarray}
We also compared to a dipole parametrization ($\sigma=0.0584$,
$\rho=0$) and to a monopole form ($\sigma=\rho=0$).

For the kaonic form factors ($M=K$), consider the parametrizations
\begin{eqnarray}
   F^{(K)}_V(t) &= &F^{(K)}_V(0)  \left[\Res_{\kst}(t) +
        \sigma_K \Res_{\kst'}(t) + \rho_K \Res_{\kst''}(t) \right]
        \frac{1}{1 + \sigma_K + \rho_K}
\nonumber \\
    F_{A}^{(K)} (t) & = & F_{A}^{(K)}(0) \Res_{K_1(1270)} (t)
\end{eqnarray}
(where $\kst = \kst(892)$, $\kst' = \kst(1410)$ and $\kst'' =
\kst(1680)$).
In \cite{Dec93} we used the monopole parametrization
$\sigma_K = \rho_K = 0$ of the vector form factor and
constant widths
for both the $K^\star$ and the $K_1$.

Flavour symmetry implies the following relations for the
form factors at $t = 0$:
\begin{eqnarray} 
   F_{A}^{(K)} (0) & = & \frac{m_K}{m_\pi} F^{(\pi)}_A(0) = + 0.0410
   \nonumber \\
   F_{V}^{(K)} (0) & = & \frac{m_K}{m_\pi} F^{(\pi)}_V(0) = + 0.0955
\end{eqnarray}
The measurement of $K\to l \nu_l \gamma$, on the other hand
\cite{RPP92}, gives
\begin{eqnarray}
   F_{V}^{(K)} + F_A^{(K)} & = & + 0.148 \pm 0.010
\nonumber \\[0.5ex]
   F_{V}^{(K)} - F_A^{(K)} & \in & [- 0.3 , 2.2]
\end{eqnarray}
Taking the vector form factor from flavour symmetry and the
anomaly and calculating the axial one from the measured sum and the
value for $F_V^{(K)}(0)$, this results in
\begin{eqnarray} \label{eqn2133}
   F_{A}^{(K)} (0) & = & \frac{m_K}{m_\pi} F_A^{(K)}(0) = + 0.0525 \pm 0.010
   \nonumber \\
   F_{V}^{(K)} (0) & = & \frac{m_K}{m_\pi} F_V^{(K)}(0) = + 0.0955
\end{eqnarray}
These are the values we will actually use.
%
\section{General Considerations on the Treatment of the Virtual Corrections}
\label{sec31}
The momentum dependence of the $W$ propagator
\be
    \frac{1}{m_W^2}\frac{m_W^2}{m_W^2 - q^2}
\ee
is determined by the familiar Feynman cut-off function
$ m_W^2/(m_W^2 - q^2)$.
Thus we can as well use a local interaction with an UV cut-off
equal to $m_W$ in the calculation of the virtual corrections.
In fact, it has been shown by Sirlin \cite{Sir78}
that the radiative
corrections of the order $G_F \alpha$ calculated within the full
standard model with a single Higgs doublet are equal to the photonic
corrections calculated with the local $V-A$ interaction and a cut-off
equal to $m_Z$, except for very small contributions of the order
$\alpha_s G_F \alpha$.
While the photonic corrections are identical to
those computed in the local $V-A$ theory with an effective cut-off
equal to $m_W$, the non-photonic corrections lead to an additional
contribution which depends on $\ln (m_Z/m_W)$ and $\theta_W$. In the
simplest electroweak model with a single Higgs doublet, where $\cos
\theta_W = m_W/m_Z$, the photonic and the non-photonic corrections
combine to give a result which is identical to the photonic correction
obtained in the local theory with the cut-off equal to $m_Z$.
As we will show, the actual value of the cut-off does not matter for our
results on $R_{\tau/\pi}$, because it cancels in this ratio.

Thus we will use a local $V-A$ interaction and dimensional
regularization, ie.\ the loop integrals are evaluated in $4 -\epsilon$
rather than in 4 space-time dimensions.
We can translate our results into a momentum space cut-off by the
replacement
\be \label{eqn34}
   \Delta - \ln \frac{m^2}{\mu^2} \longrightarrow \ln \frac%
   {m_Z^2}{m^2}
\ee
where
\be
   \Delta := \frac{2}{\epsilon} - \gamma_{Euler} + \ln 4 \pi
\ee
\label{sec32}
The virtual corrections fall into three classes, see \fig{figc1}.
The first
class includes only a single diagram in which all couplings are known,
and the third class is identical for $\tau\to\pi\nu_\tau$ and
$\pi\to\mu\nu_\mu$, so these diagrams drop out in the ratio
$R_{\tau/\pi}$.
Therefore the most important class for the calculation of
$R_{\tau/\pi}$ is the second one.

Now we want to separate the integration over the
momentum of the virtual photon into two regions with small and large
$k_E^2$, respectively \cite{Sir72}.
We achieve this by  splitting  the photon
propagator:
\begin{eqnarray}
   \frac{1}{k^2 - \lambda^2} & = &
   \underbrace{\frac{1}{k^2 - \lambda^2}\frac{\mu_{cut2}^2}
   {\mu_{cut}^2 - k^2}}_{\mbox{``long distance''}}
   + \underbrace{\frac{1}{k^2 - \mu_{cut}^2}}_{\mbox{``short distance''}}
\label{eqn36}
\end{eqnarray}
Obviously the
first part is important only for $|k^2| \lo \mu_{cut}^2$ and the
second part only for $|k^2| \go \mu_{cut}^2$, so indeed they correspond
to long and short distances, respectively.
And so the photon propagator is divided  into two parts,
a regulated photon propagator with an effective cut-off
$\mu_{cut}$ and a massive photon propagator with mass $\mu_{cut}$.
The scale $\mu_{cut}$ separates long and
short wavelengths and should be of the order $\mu_{cut} \sim O(1
\unit{GeV})$.

\begin{figure}
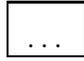

   \caption{Short distance corrections}
   \label{figc2}
\begin{center}
\fbox{\str \ \dots \ }
  \end{center}
\end{figure}

\begin{figure}
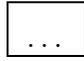

   \caption{Long distance corrections}
   \label{figc3}
\begin{center}
\fbox{\str \ \dots \ }

  \end{center}
\end{figure}

A photon with short
wavelength ($|k^2| \go \mu_{cut}^2$) resolves the quarks in the pion.
It interacts with the quarks which come into being in the
initial process $\tau\to \nu_\tau \bar{u} d$, which subsesquently
hadronize to form the pion (see \fig{figc2}). A photon with long
wavelength ($|k^2| \lo \mu_{cut}^2$), on the other hand, has a small
resolution and interacts with the pion as a whole or perhaps with some
hadronic resonances (see \fig{figc3}). So according to \eqn{eqn36}
we have to calculate the
short distance diagrams such as in \fig{figc2} with a massive photon
propagator and the long distance diagrams such as in \fig{figc3} with a
regulated photon propagator.

In the next section we will start with the calculation of the
corrections in a point pion model.
This will give a first estimate, and it is part of the complete
calculation.
%
\section{Point Meson Contribution}
%
\label{sec33}

We will now calculate the corrections to $\tau\to\pi\nu_\tau$ and
$\pi\to\mu\nu_\mu$ in a model with an effective point pion field,
using a generalized
photon propagator with a mass $m^2$
\be
   \frac{1}{k^2 - \lambda^2} \to \frac{1}{k^2 - m^2}
\ee
($\lambda$ denotes a small IR regulator mass, which in the end of the
calculation is put equal to zero, whereas $m$ denotes any finite
mass.)
It will
become clear below why the consideration of such a generalized photon
propagator is useful, see Eqns.~(\ref{eqn388})--(\ref{eqn391}).

For the case of the pion decay (and for $m^2 = \lambda^2$), these
corrections have been calculated in
\cite{Kin59}.
%
Our
calculation, however, differs in some technical details from that in
 \cite{Kin59}. First we use dimensional rather that cut-off
regularization. Second Kinoshita replaces
the vector-minus-axial vector current interaction
\be \label{eqnvma}
   G_F \cos \theta_C f_\pi [\bar{\Psi}_l \gamma^\mu (1 - \gamma_5)
   \Psi_\nu] (i \partial_\mu - e A_\mu ) \Phi_\pi
\ee
by the scalar-minus-pseudoscalar current interaction
\be \label{eqnsmp}
   G_F \cos \theta_C f_\pi m_l^{(0)}
   [\bar{\Psi}_l  (1 - \gamma_5) \Psi_\nu] \Phi_\pi
\ee
where $m_l^{(0)}$ denotes the bare lepton mass. We have performed the
calculation both using \eqn{eqnvma} and using \eqn{eqnsmp} and we have
checked that they give identical results. Here we present the results
using the $V-A$ form of \eqn{eqnvma}.

\begin{figure}
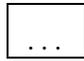

   \caption{Point pion diagrams}
   \label{figc4}
\begin{center}
\fbox{\str \ \dots \ }
  \end{center}
\end{figure}

There are six Feynman diagrams for the virtual
corrections to $\tau\to\pi\nu_\tau$ , see \fig{figc4}, and of course
there are six similar diagrams for $\pi\to\mu\nu_\mu$.
The last diagram $\delta \mcal_T$ actually vanishes after pion mass
renormalization.

The ratios of  the virtual correction amplitudes
over the Born amplitudes for both the tau and the
meson decay can be expressed by the same functions of $m_l^2$ and $m_M^2$,
if we use a general lepton mass $m_l$ which denotes
$m_\tau$, $m_\mu$ or $m_e$, respectively.
The result is
\begin{eqnarray}
   \frac{\delta \mcal_1}{\mcal_0} (m_l^2,m_M^2,m^2)
   & = & \frac{\alpha}{4 \pi}
   \Bigg\{2 B_0^{M} - B_1^M + 2 m_l^2 C_0 - 2 m_M^2 C_1 \Bigg\}
\nonumber \\
\nonumber \\
   \frac{\delta \mcal_2}{\mcal_0}  (m_l^2,m_M^2,m^2)
   & = & \frac{\alpha}{4 \pi}
   \Bigg\{ 1 - 4 B_0^l - 2 B_1^l \Bigg\}
\nonumber \\
\nonumber \\
   \frac{\delta \mcal_3}{\mcal_0} (m_l^2,m_M^2,m^2)
    & = & \frac{\alpha}{4 \pi}
   \Bigg\{- B_0^M + B_1^M \Bigg\}
\nonumber \\
\nonumber \\
   \frac{\delta \mcal_4}{\mcal_0} (m_l^2,m_M^2,m^2)
    & = & \frac{\alpha}{4 \pi}
   \Bigg\{ \frac{1}{2} B_0^M - B_1^M + m_M^2 [{B'_0}^{M} -
   {B'_1}^M ] \Bigg\}
\nonumber \\
\nonumber \\
   \frac{\delta \mcal_5}{\mcal_0}  (m_l^2,m_M^2,m^2)
   & = & \frac{\alpha}{4 \pi}
   \Bigg\{ \frac{1}{2} + B_1^l + m_l^2 [4 {B'_0}^l + 2 {B'_1}^l ] \Bigg\}
\end{eqnarray}
where the $B_i^l$, $B_i^\pi$ and $C_i$ are given in terms of the standard
n-point functions $B$ and $C$:
\begin{eqnarray}
   C_i & = & C_i(m_M^2,m_l^2,0,m_M,m,m_l)
\nonumber \\
   B_i^l & = & B_i(m_l^2,m_l,m)
\nonumber \\
   B_i^M & = & B_i(m_M^2,m_M,m)
\end{eqnarray}
(see \app{appa} for our conventions). Thus
\pagebreak[3]
\begin{eqnarray}
   \frac{\delta \mcal_{PML}}{\mcal_0}(m_l^2,m_M^2,m^2)
   & := & \sum_{i=1}^{6} \frac{\delta \mcal_i}{\mcal_0}
\nonumber \\
    & = & \frac{\alpha}{4 \pi} \Bigg\{ \frac{3}{2} + \frac{3}{2} B_0^M
   - B_1^M + m_M^2 [{B'_0}^M - {B'_1}^M]
   - 4 B_0^l - B_1^l
\nonumber \\
& & \quad + {} m_l^2[4 {B'_0}^l + 2 {B'_1}^l ]
   + 2 m_l^2 C_0 - 2 m_M^2 C_1 \Bigg\}
\end{eqnarray}
(``PML'' denotes ``point meson loops''.) This is the form which will
be used in the calculation of the complete radiative correction in
\sect{sec38}.

To obtain the radiative correction within a model with a pointlike
pion, we now assume $m^2 = \lambda^2$ and
calculate the $n$-point functions. The result is
\begin{eqnarray}
   \frac{\delta \mcal_{PML}}{\mcal_0}(m_l^2,m_M^2,\lambda^2)
   & = & \frac{\alpha}{4 \pi} \Bigg\{ - \frac{3}{2} \Delta
   + \frac{3}{2} \ln \frac{m_M^2}{\mu^2} - 4
   - 4 \left[ \frac{1+r_l^2}{1-r_l^2} \ln r_l + 1 \right]
   \ln \frac{\lambda}{m_M}
\nonumber \\ & & \quad {}
   + 2 \frac{1+r_l^2}{1-r_l^2} (\ln r_l)^2
   + \left( 5 - \frac{4 r_l^2}{1-r_l^2}\right) \ln r_l \Bigg\}
\end{eqnarray}
where
\be
   r_l := \frac{m_l}{m_M}
\ee
and $\mu$ is the mass scale of dimensional regularization.

By interference with the Born amplitude, the virtual correction of the
decay rate is of course
\be
   \frac{\delta\Gamma_{PML}}{\Gamma_0} = 2 \frac{\delta \mcal_{PML}}
   {\mcal_0}
\ee

To these virtual corrections the integrated decay rate for internal
bremsstrahlung must be added. We divide this rate
into ``soft'' ($E_\gamma \leq E_0$) and ``hard'' ($E_\gamma \geq
E_0$) corrections,
\be
   \frac{\delta \Gamma_{IB} (\tau\to M \nu_\tau\gamma)}
   {\Gamma_0(\tau\to M\nu_\tau)}  =
   \frac{\delta \Gamma_{soft}^\tau}{\Gamma_0^\tau} +
   \frac{\delta \Gamma_{hard}^\tau}{\Gamma_0^\tau}
\ee
where $E_0$ is assumed to be small, ie.\ $x_0 := (m_\tau/2) E_0 \ll 1$.
For the $\tau$ decay the results are
\begin{eqnarray}
   \frac{\delta \Gamma_{soft}^\tau(x \leq x_0)}{\Gamma_0^\tau} & = &
   \frac{\alpha}{2 \pi} \Bigg\{ 4 \left( \frac{1 + r_M^2}{1-r_M^2}
   \ln r_M + 1 \right) \left(\ln \frac{\lambda}{m_M} - \ln x_0
   \right)
   + 2 \frac{1+r_M^2}{1-r_M^2} (\ln r_M)^2
\nonumber \\
\nonumber \\
 & & \quad {} + \frac{2 - 6 r_M^2}
   {1-r_M^2} \ln r_M + 2 - \frac{1}{3} \pi^2
   + 2 \frac{1+r_M^2}{1-r_M^2} \Li (r_M^2) - \frac{2}{3}
   \frac{r_M^2}{1-r_M^2} \pi^2
\nonumber \\
\nonumber \\ & & \quad {}
   + 4 \frac{1+r_M^2}{1-r_M^2}
   \ln r_M \ln (1-r_M^2)
   + O(x_0) \Bigg\}
\end{eqnarray}
and
\begin{eqnarray}
   \frac{\delta \Gamma_{hard}^\tau(x \geq x_0)}{\Gamma_0^\tau} & = &
   \frac{\alpha}{2 \pi} \Bigg\{ 4 \left( \frac{1+r_M^2}{1-r_M^2}
   \ln r_M + 1 \right) \ln x_0 + \frac{25}{4} - \frac{1}{3} \pi^2
   + \frac{4 - 2 r_M^2+r_M^4}{(1-r_M^2)^2} \ln r_M
\nonumber \\
\nonumber \\
 & & \quad {}
   + \left(\frac{3}{2} - \frac{2}{3} \pi^2 \right) \frac{r_M^2}
   {1-r_M^2} - 4 \ln(1-r_M^2)
   + 2 \frac{1+r_M^2}{1-r_M^2} \Li(r_M^2)
\nonumber \\
\nonumber \\ & & \quad {}
     + O(x_0) \Bigg\}
\end{eqnarray}
where
\be
   r_M := \frac{m_M}{m_\tau}
\ee

For the meson decay the sum of soft and hard bremsstrahlung is
\be
   \frac{\delta \Gamma_{IB} (M\to l \nu_l\gamma)}
   {\Gamma_0(M\to l\nu_l)}  =
   \frac{\delta \Gamma_{soft}^M(x \leq x_0)}{\Gamma_0^M} +
   \frac{\delta \Gamma_{hard}^M(x \geq x_0)}{\Gamma_0^M}
\ee
\begin{eqnarray}
 & = & \frac{\alpha}{2 \pi} \Bigg\{ 4 \left(\frac{1+r_l^2}
   {1-r_l^2} \ln r_l + 1 \right) \ln \frac{\lambda}{m_M}
   - 2 \frac{1+r_l^2}{1-r_l^2} (\ln r_l)^2
\nonumber \\
\nonumber \\ & & \quad {}
      - 4 \left( \frac{1+r_l^2}{1-r_l^2}
   \ln r_l + 1 \right) \ln (1-r_l^2)
   + \frac{1-6r_l^2 + 2 r_l^4}{(1-r_l^2)^2} \ln r_l
\nonumber \\
\nonumber \\ & & \quad {}
      - 4 \frac{1+r_l^2}{1-r_l^2} \Li(1-r_l^2) + \frac{27}{4}
   - \frac{3}{2} \frac{r_l^2}{1-r_l^2} \Bigg\}
\end{eqnarray}
In these formulae, $\Li$ denotes the dilogarithmic  function
\be
   \Li(x) = - \int_0^x dt \, \frac{\ln(1-t)}{t}
\ee

Now we add up virtual, soft and hard photonic corrections as obtained
in the point meson model.   Writing
the radiatively corrected rates as
\be
   \frac{\Gamma(\mbox{channel})}{\Gamma^0(\mbox{channel})} =
     1 + \frac{\delta\Gamma}{\Gamma^0}\Big(\mbox{channel}\Big)
\ee
we obtain
\be \label{eqn325}
   \frac{\delta\Gamma}{\Gamma^0}\Big(M \to l \nu_l (\gamma)\Big) =
   \frac{\alpha}{2 \pi} \Bigg\{
   - \frac{3}{2} \Delta
   + \frac{3}{2} \ln \frac{m_M^2}{\mu^2} + 6 \ln r_l + \frac{11}{4}
   - \frac{2}{3} \pi^2
   + f(r_l) \Bigg\}
\label{eqnext16}
\ee 
and
\be \label{eqnext17} \label{eqn326}
   \frac{\delta\Gamma}{\Gamma^0}\Big( \tau \to M \nu_\tau (\gamma) \Big) =
   \frac{\alpha}{2 \pi} \Bigg\{
      - \frac{3}{2} \Delta
   + \frac{3}{2} \ln \frac{m_\tau^2}{\mu^2} + \frac{17}{4}
    - \frac{2}{3} \pi^2 + g(r_M) \Bigg\}
\ee 
where
\begin{eqnarray}
   f(r_l) & = & 4 \left(\frac{1+r_l^2}{1-r_l^2} \ln r_l
   - 1 \right) \ln(1-r_l^2)
   - \frac{r_l^2 (8 - 5r_l^2)}{(1-r_l^2)^2}\ln r_l
\nonumber \\ \nonumber \\
   & & + 4 \frac{1 + r_l^2}{1-r_l^2} \mbox{Li}_2(r_l^2)
   - \frac{r_l^2}{1-r_l^2}
   \left( \frac{3}{2} + \frac{4}{3} \pi^2 \right)
\end{eqnarray}
and
\begin{eqnarray}
   g(r_M) & = & 4 \left( \frac{1+r_M^2}{1-r_M^2} \ln r_M - 1 \right)
                                       \ln(1-r_M^2)
   - \frac{r_M^2(2-5r_M^2)}{(1-r_M^2)^2} \ln r_M
\nonumber \\ \nonumber \\
   &  & + 4 \frac{1+r_M^2}{1-r_M^2} \mbox{Li}_2(r_M^2)
   + \left( \frac{3}{2} - \frac{4}{3} \pi^2 \right) \frac{r_M^2}{1-r_M^2}
\end{eqnarray}

We have written the corrections in such way that
$f(0) = g(0) = 0$, so from Eqn.\ (\ref{eqnext16}) we get the well-known
lepton mass singularity of the total radiative correction to pion decay as
$3 \alpha / \pi \ln (m_l)$, whereas in the radiative correction
to the tau decay the meson mass singularities cancel according to Eqn.
(\ref{eqnext17}).
Our results are a nice application of the Kinoshita-Lee-Naunenberg
theorem, which states that mass singularities cancel in inclusive
decay rates \cite{Kin59,Que92}. In the case of the pion decay the Born
amplitude is proportional to $m_l$ and therefore a logarithm $\ln m_l$
is allowed in the radiative correction, while for the tau decay the
Born amplitude is not proportional to $m_\pi$, and therefore a
logarithm $\ln m_\pi$ is forbidden in the radiative correction.

The ultra-violet divergences of the corrections to the tau and
the meson decays are equal and they cancel in the ratio $R_{\tau/M}$
defined by
\be
   R_{\tau/M} = \frac{\Gamma(\tau \to M \nu_\tau (\gamma))}
   {\Gamma(M \to \mu \nu_\mu (\gamma))} =
   \frac{1}{2}
   \frac{m_\tau^3}{m_M m_\mu^2}
   \frac{\left(1 - m_M^2 / m_\tau^2\right)^2}
   {\left(1 - m_\mu^2 / m_M^2 \right)^2} \Bigg(1 + \delta R_{\tau/M} \Bigg)
\ee
with the finite radiative correction
\be
  \Big(\delta R_{\tau/M} \Big)_{PM}
   = \frac{\alpha}{2 \pi}
   \left\{ \frac{3}{2} \ln \frac{m_\tau^2}{m_M^2}
   - 6 \ln \frac{m_\mu}{m_M} + \frac{3}{2} + g\left(\frac{m_M}
   {m_\tau}\right) - f\left(\frac{m_\mu}{m_M}\right) \right\}
\ee
So within this model with an effective pointlike meson (``PM''),
we end up with the results~\cite{Dec93b}
\begin{eqnarray} \left(
     \delta R_{\tau/\pi} \right)_{PM} & = & + 1.05 \%
\nonumber \\ \left(
     \delta R_{\tau/K}\right)_{PM} & = & + 1.67 \%
\end{eqnarray}
Note that this result differs from the Marciano-Sirlin estimate quoted
in the introduction,
\begin{eqnarray}
   \delta R_{\tau/\pi}^{M.S.} & = & - 0.98 \%
\nonumber \\[0.5ex]
   \delta R_{\tau/K}^{M.S.} & = & - 0.53 \%
\end{eqnarray}
This is mainly due to the fact that the point meson and
the short distance corrections do not have the same UV divergences.

{}From \eqn{eqnext16} the radiative
correction to the ratio of the electronic and the muonic decay modes
of the pion
\be
   R_{e/\mu} = \frac{\Gamma(\pi \to e \nu_e (\gamma))}
   {\Gamma(\pi \to \mu \nu_\mu (\gamma))} =
   \frac{m_e^2}{m_\mu^2} \left(\frac{m_\pi^2 - m_e^2}{m_\pi^2 - m_\mu^2}
   \right)^2 \Bigg(1 + \delta R_{e/\mu}\Bigg)
\ee
can also be calculated in the same way and we obtain
\be
   \Big(  \delta R_{e/\mu}  \Big)_{PM}
   = \frac{\alpha}{2 \pi}
   \left\{6 \ln \frac{m_e}{m_\mu} + f\left(\frac{m_e}{m_\pi}\right)
    - f\left(\frac{m_\mu}{m_\pi}\right) \right\}
  =  - 3.93 \%
\ee
a result derived long ago by Kinoshita \cite{Kin59}.
%
\section{The Leading Short Distance Logarithm}
\label{sec34}
\label{sec341}
In Ref.~\cite{Sir82} the author shows that
for any semileptonic weak process,
the leading $O(\alpha)$ correction in the limit of a large $Z$~boson
mass, $m_Z^2 \to \infty$, is
\be \label{eqntheorem}
    {\mcal_0} \to \left[ 1 + \frac{\alpha}{\pi}\ln \frac{m_Z}{\mu}
     \right] \mcal_0
\ee
where $\mcal_0$ denotes the Born amplitude and $\mu$
is an unspecified mass scale characteristic of the process.

\begin{figure}
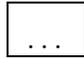

   \caption{Short distance diagrams}
   \label{figc6}
\begin{center}
\fbox{\str \ \dots \ }
  \end{center}
\end{figure}

We will now rederive this result and then show that we are able to fix
the scale $\mu$ in terms of the lepton mass $m_l$ and the scale
$\mu_{cut}$.
Consider the short distance corrections to the amplitude
$\acal_0$ for the initial weak process $\tau \to \nu_\tau \bar{u} d$
\be \label{eqn335}
   \acal_0 = -i \frac{G_F \cos \theta_C}{\sqrt{2}}
    [\bar{u}_d \gamma^\mu \gm u_u] [\bar{u}_\nu \gamma_\mu \gm u_\tau]
\ee
The six
Feynman diagrams which give the radiative corrections to this short
distance amplitude are shown in \fig{figc6}.
In order to find the term
of the order $O(\alpha \ln m_Z)$, we have to calculate the UV
divergence of the short distance correction
\be
  \delta \acal = \delta \acal_a + \delta \acal_b + \dots + \delta \acal_f
\ee
The calculation  can be simplified by using
the Landau gauge \cite{Geo84}. In this gauge the amplitudes
corresponding to external line renormalization, $\delta \acal_a \dots
\delta \acal_c$, are UV finite, as are
$\delta \acal_e$ and $\delta \acal_f$, where the photon loop connects an
ingoing with an outgoing fermion line.
In the Landau gauge, the only UV divergent amplitude is $\delta \acal_d$:
\begin{eqnarray} \label{eqn338}
   \delta \acal_d & = & e^2 Q_u Q_\tau \frac{G_F}{\sqrt{2}} \mu^{D-4}
   \int \frac{d^D k}{(2 \pi)^D} \frac{
   [\bar{u}_d \gamma^\mu \gm  \strich{k} \gamma^\alpha u_u]
   [\bar{u}_\mu \gamma_\mu \gm \strich{k} \gamma^\beta \bar{u}_\tau]}%
   {(k^2)^3}
    \left[ g_{\alpha\beta}- \frac{k_\alpha k_\beta}{k^2}\right]
   + \dots
\nonumber \\
  & = & \frac{\alpha}{2 \pi} \Delta \, \acal_0 + \dots
\end{eqnarray}
where the dots indicate terms which are UV finite.

Now replacing $\Delta$ by the UV cut-off
$\ln m_Z^2$, we obtain
\be
  \delta \acal = \frac{\alpha}{ \pi} \ln \frac{m_Z}{\mu}\, \acal_0 + \dots
\ee
where $\mu$ is some unspecified characteristic
mass scale which must be introduced by
dimensional arguments. This is Sirlin's result.

If we replace the photon propagator $1/k^2$ by $1/(k^2 - \mu_{cut}^2)$
and if we do not neglect the lepton mass, we obtain
\be
  \delta \acal = \frac{\alpha}{\pi} \frac{1}{m_l^2 - \mu_{cut}^2}
   \left(m_l^2 \ln \frac{m_Z}{m_l} - \mu_{cut}^2 \ln
   \frac{m_Z}{\mu_{cut}} \right) \acal_0 + \dots
\ee

We will now apply this result to the radiative correction to
$R_{\tau/\pi}$.
In addition to the  short distance correction which contains the above
logarithm, there are
the long distance corrections associated with
the integration over $k_E^2 =
0 \dots \mu_{cut}^2$, which have to be added.
Of course these long distance corrections have to be computed using
the effective pointlike pion (and hadronic resonances) and give rise
to other logarithms which in fact turn out to be more important in
$R_{\tau/\pi}$ that the short distance ones.

So we will now combine  the long and the short distance corrections.
First we have to discuss which scale
is a good choice for $\mu_{cut}$. The effective point meson is
a good approximation only if all momentum transfers squared are small
compared to $m_\rho^2$, and so for the long distance part one should
require $\mu_{cut}^2 \ll m_\rho^2$. The short distance part, on the other
hand, uses asymptotically free quarks which one would believe in only
for $\mu_\subs{cut}^2 > (1 \dots 2  \unit{GeV})^2$.  A compromise
between these
non-overlapping regions would be $\mu_{cut}^2 = m_\rho^2$.

The long distance corrections
are to be integrated over $k_E^2 = 0 \dots \mu_{cut}^2$ ,
which is taken into account by using an UV cut-off
$\mu_{cut} = m_\rho$ for the
point meson results of \sect{sec33} (ie.\ by the replacement $\Delta
\to \ln (m_\rho^2/\mu^2)$ in Eqns.~(\ref{eqn325}) and (\ref{eqn326})).
Thus we obtain
\be \label{eqn344}
   \frac{\delta\Gamma}{\Gamma^0}\Big(\pi \to l \nu_l (\gamma)\Big) =
   \overbrace{
   \underbrace{\frac{2 \alpha}{ \pi} \ln \frac{m_Z}{m_\rho}}_{\D2.22\%}}^%
   {\mbox{\small short dist.}}
   + \overbrace{\underbrace{\frac{\alpha}{2 \pi} \Bigg\{
  \frac{3}{2} \ln \frac{m_\pi^2}{m_{\rho}^2} + 6 \ln r_l + \frac{11}{4}
   - \frac{2}{3} \pi^2
   + f(r_l) \Bigg\}}_{\D -1.02 \%}}^{\mbox{\small long dist.}}
\ee 
and
\begin{eqnarray}
 \label{eqn345}
   \frac{\delta\Gamma}{\Gamma^0}\Big( \tau \to \pi \nu_\tau (\gamma)
  \Big) & = &
   \overbrace{\underbrace{
  \frac{2 \alpha}{\pi} \frac{1}{m_\tau^2 - m_\rho^2}
   \left(m_\tau^2 \ln \frac{m_Z}{m_\tau} - m_\rho^2 \ln
   \frac{m_Z}{m_\rho} \right)
   }_{\D 1.74 \%}}^%
   {\mbox{\small short dist.}}
\nonumber \\[0.3ex]  & & \qquad \qquad \qquad
 + \quad \overbrace{\underbrace{\frac{\alpha}{2 \pi} \Bigg\{
   \frac{3}{2} \ln \frac{m_\tau^2}{m_{\rho}^2} + \frac{17}{4}
    - \frac{2}{3} \pi^2 + g(r_\pi) \Bigg\}}_{\D 0.03 \%}}^
   {\mbox{\small long dist.}}
\end{eqnarray} 
This modifies the point meson results in the following way:
\begin{eqnarray} \label{eqnpointplussd}
   \delta R_{\tau/\pi} & = &\Big(\delta R_{\tau/\pi}\Big)_{PM}
   + \underbrace{
   \frac{2 \alpha}{\pi} \frac{m_\tau^2}{m_\tau^2 - m_\rho^2}
   \ln \frac{m_\rho}{m_\tau}
   }_{\D -0.48\%}
\nonumber \\
   & = & 0.57 \%
\end{eqnarray}

Similarly
\begin{eqnarray}
   \delta R_{\tau/K} & = &\Big(\delta R_{\tau/\pi}\Big)_{PM}
   + \underbrace{
   \frac{2 \alpha}{\pi} \frac{m_\tau^2}{m_\tau^2 - m_{K^\star}}
   \ln \frac{m_{K^\star}}{m_\tau}
   }_{\D -0.43\%}
\nonumber \\
   & = & 1.24 \%
\end{eqnarray}

The prediction for $\delta R_{\tau/\pi}$ obtained in this way by matching the
leading short distance logarithms with the effective point pion
correction, $\delta R_{\tau/\pi} = 0.57\%$, differs strongly from the
first Marciano-Sirlin estimate \cite{Mar88}, $\delta R_{\tau/\pi} = - 0.98
\% $, which
is based on the short distance logarithms only.
Note, however, that it is numerically close to the predictions in
\cite{Mar92,Mar93}, see Eqns.~(\ref{eqnmarciano},\ref{eqnms2}) above.
The estimate in \cite{Mar92} (and up to very small additional
corrections, the value in \cite{Mar93} as well)
includes long and short distance corrections as in \eqn{eqn344} for
the pion decay. For the tau decay, the authors only include the short
distance correction, estimated using \eqn{eqntheorem} with
$\mu=m_\tau$, ie.\ they estimate the  distance correction to be $(2 \alpha /
\pi) \ln(m_Z/m_\tau) = 1.83 \%$, which has to be compared to the $1.74
\%$ calculated in \eqn{eqn345}.

And so,
what is really missing in
\cite{Mar92,Mar93}, as compared to our estimate in
\eqn{eqnpointplussd}, are the long distance
corrections to the tau decay.
In \cite{Mar92,Mar93} the authors only estimated their possible size,
resulting in a $\pm 1 \%$ uncertainty.
As we have calculated, the long distance corrections in the model with
an effective point meson
happen to be extremely small ($0.03 \%$), which explains why our
estimate in \eqn{eqnpointplussd} and the estimates in
\cite{Mar92,Mar93} are quite similar numerically.

However, these estimates still cannot be considered as safe for the
following reasons:
\begin{enumerate}
\item
The value $\mu_{cut} = m_\rho$ is too large for the point meson
approximation in the long distance part and too small for the
assumption of almost free quarks in the short distance part. Therefore
the range of validity in the long distance part should be extended to
$1 \dots 2 \unit{GeV}$ by including vector meson resonances and
then $\mu_{cut} = 1 \dots 2 \unit{GeV}$ will be a good value to
use.
Indeed we will show below that the vector meson effects in the loops
change the final result considerably.
\item
For the long distance corrections to the tau decay, the UV cut-off
$\mu_{cut}$ is smaller than  $m_\tau$.
Therefore terms
proportional to $m_\tau^2/ \mu_{cut}^2$ are missing in \eqn{eqn345}.
\item
The large scale $m_Z$ appearing in the short distance logarithm
cancels in the ratio $R_{\tau/\pi}$, leaving logarithms of comparable
scales ($\ln m_\rho$ and $\ln m_\tau$). Therefore it is not obvious
that
non-logarithmic contributions in the short distance correction can
be neglected.
\end{enumerate}

So in the following sections we will improve
the calculation of the long distance
corrections by properly using the regulated photon propagator of
\eqn{eqn36} and by
including vector meson resonances, and that of
of the short
distance corrections by going beyond the leading logarithm.
%
\section{Short Distance Beyond Leading Logarithm}%
%
\label{sec342}

In this section we will present a complete calculation of the short
distance corrections beyond the logarithm which is leading in the
limit $m_Z^2 \to \infty$. This leading logarithmic contribution to the
correction $\delta \acal$ of the short distance amplitude
$\acal_0(\tau\to\nu_\tau\bar{u}d)$ could be written as a factor times
the Born amplitude, $\delta \acal = C \times \acal_0 + \dots$. Then
the correction $\delta \mcal$ to the exclusive rate
$\mcal_0(\tau\to\pi\nu_\tau)$ involves the same logarithm, $\delta
\mcal = C \times \mcal_0 + \dots$. The situation, however, is more
involved for the complete result for $\delta \acal$. Firstly, in
general it can not be written as a number times $\acal_0$, but rather
it involves other operators, and secondly $\delta \acal$ will depend
on the relative momentum of the two quarks. The first problem can be
solved by projection on the $J^P = 0^-$ state of the two quarks, and
the second problem by use of the parton model.

\begin{figure}
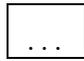

\caption{Decay $\tau\to\pi\nu_\tau$ via an intermediate
quark-antiquark state.  The bubble on the left hand side is to be
replaced by the short
distance diagrams $\acal_0$ and $\delta \acal_i$, cf.\ Fig.~%
\protect\ref{figc6}}
\label{figw}
\begin{center}
\fbox{\str \ \dots \ }
\end{center}
\end{figure}

Let the short distance amplitude for the decay
\be
   \tau(s) \to \nu_\tau(q) \bar{u}(\frac{1}{2} p -l) d(\frac{1}{2} p +
l)
\ee
 of the tau into  neutrino and an
quark-antiquark pair with relative momentum $2 l$
be
given by $\acal(l)$ (and similarly for the decay into $\bar{u}s$, for
the time being we consider the case of non-strange decay into the pion
only).
In a frame, where the pion is moving with infinite
momentum, $l$ is proportional to $p$,
\be
    l = \frac{u}{2} p
\ee
and $\acal = \acal(u)$.
Then the amplitude of the exclusive tau decay $\tau\to\pi\nu_\tau$ is
(see \fig{figw})
\be \label{eqn348}
   \mcal(\tau^- \to \pi^-\nu_\tau) = \tcal(\acal)
\ee
where
\be
   \tcal(\acal) =
   - i \frac{3 \sqrt{2}}{8} \int_{-1}^{+1} du \, \frac{\Phi_\pi(u)}
   {m_q} (-) \sum\left[ \acal(u) \bar{u}_u\left(\frac{1-u}{2} p \right)
   \gamma_5 u_d\left(\frac{1+u}{2}p \right) \right]
\ee
Here $\Phi_\pi(u) = \Phi_\pi(-u)$  is a symmetric parton distribution
function
(The numerical factor is just conventional. Our
treatment follows closely that of Refs.\
\cite{Koe74,Kue83}).
The sum $ \sum$ is over the spins of the quarks.
For the quark mass $m_q$ we assume isospin (and in the case of the
kaon, $SU(3)$-flavour) symmetry,
\be
  m_q = m_u = m_d = m_s
\ee
The limit $m_q \to 0$ is implied. Note that all relevant operators in
$\acal$ involve an odd number of Dirac matrices between the
$\bar{u}_d$ and the $u_u$ Dirac spinors, so the sum is proportional to
$m_q$, and this $m_q$ cancel the $m_q$ in the denominator.

In the case of the  Born amplitude $\mcal_0$, $\acal_0$ is
\be
   \acal_0 = -i \frac{G_F \cos \theta_c}{\sqrt{2}}
    [\bar{u}_d \gamma^\mu \gm u_u] [\bar{u}_\nu \gamma_\mu \gm u_\tau]
\ee
and so
\begin{eqnarray}
\lefteqn{   \mcal_0(\tau^- \to \pi^-\nu_\tau)  =  \tcal(\acal_0)}
\nonumber \\
 & = & - \frac{3}{8} G_F \cos \theta_c \int_{-1}^{+1} du \,
   \frac{\Phi_\pi(u)}{m_q} (-) \tr \left[ \gamma^\mu \gm
   \left(\frac{1-u}{2} \strich{p} + m_q\right) \gamma_5
   \left(\frac{1+u}{2} \strich{p} + m_q\right) \right]
\nonumber \\[0.5ex] & & \qquad \qquad \times
   \Big[\bar{u}_\nu \gamma_\mu \gm u_\tau\Big]
\nonumber \\[0.5ex] & = &
   - \frac{3}{2} G_F \cos \theta_c \int_{-1}^{+1} du \Phi_\pi(u)
   \Big[\bar{u}_\nu \strich{p} \gm u_\tau\Big]
\nonumber \\[0.5ex] & = &
   - G_F \cos \theta_c f_\pi
   \Big[\bar{u}_\nu \strich{p} \gm u_\tau\Big]
\end{eqnarray}
where the last line follows from the definition of $f_\pi$. So we have
rederived the sum rule \cite{Bra70}
\be \label{eqnsum}
   \frac{3}{2} \int_{-1}^{+1} du \, \phi_\pi(u)
   = f_\pi
\ee

After these preparations,
we can now calculate the full short distance
corrections $\delta \acal_i$ for $i
= a \dots f$ given by the Feynman diagrams in \fig{figc6}, using
a massive photon propagator with $1/k^2 \to 1/(k^2 - \mu_{cut}^2)$,
Feynman gauge, and neglecting the quark masses.
According to \eqn{eqn348}, the corrections $\delta \acal_i$ induce
corrections $\delta \mcal_i(\tau\to M \nu_\tau)$ to the rate for the
tau decay
\be
   \delta \mcal_i(\tau\to M \nu_\tau) = \tcal(\delta \acal_i) \qquad \qquad
   (i = a \dots f)
\ee
(and analogously for the pion decay). The results depend on the
unknown distribution function $\Phi_M(u)$ and are given in the form of
integrals over $u$
\be
   \frac{\delta \mcal_i}{\mcal_0}(m_l^2, m_M^2, \mu_{cut}^2)
   = \frac{3}{2f_M} \int_{-1}^{+1} du \,
   \phi_M(u) w_i(u,m_l^2,m_M^2,\mu_{cut}^2)
\ee
with weight functions $w_i(u)$. Define the total short distance
correction amplitude
\be
   \frac{\delta \mcal_{sd}}{\mcal_0} (m_l^2, m_M^2, \mu_{cut}^2)
 = \sum_{i=a}^{f} \frac{\delta \mcal_i}{\mcal_0}(m_l^2, m_M^2, \mu_{cut}^2)
\ee
and the sum of the weight functions
\be
   w(u,m_l^2,m_M^2,\mu_{cut}^2) = \sum_{i=a}^{f} w_i(u,m_l^2,
   m_M^2,\mu_{cut}^2)
\ee
Then the short distance correction $(\delta R_{\tau/M})_\subs{sd}$ is
given by
\begin{eqnarray}
 (\delta R_{\tau/M})_\subs{sd} = 2 \left[ \frac{\delta \mcal_\subs{sd}}
  {\mcal_0}(m_\tau^2, m_m^2,\mu_{cut}^2) - \frac{\delta \mcal_\subs{sd}}
  {\mcal_0}(m_\mu^2, m_m^2,\mu_{cut}^2) \right]
\nonumber \\
  = \frac{3}{f_M} \int_{-1}^{1} du \, \phi_M(u) [w(u,m_\tau^2,
   m_M^2,\mu_{cut}^2) - w(u,m_\mu^2,m_M^2,\mu_{cut}^2)]
\end{eqnarray}
Note that $w_a$, $w_b$ and $w_e$ are identical for tau and  pion
decay, ie.\ they do not depend on the lepton mass. Using furthermore the
symmetry of the distribution function $\phi_M$ under  $u \to -u$, we
can write
\be
  (\delta R_{\tau/M})_\subs{sd} = \frac{3}{2 f_M}
   \int_{-1}^{1} du \, \phi_M(u) r_{\tau/M}(u)
\ee
with
\be
  r_{\tau/M}(u) := 2 \left[\bar{w}(u,m_\tau^2,
   m_M^2,\mu_{cut}^2) - \bar{w}(u,m_\mu^2,m_M^2,\mu_{cut}^2)\right]
\ee
where $\bar{w}$ is the symmetrized sum of the weightfunctions $w_c$,
$w_d$ and $w_f$:
\begin{eqnarray}
  \bar{w}(u,m_l^2,m_M^2,\mu_{cut}^2) & := &
  \frac{1}{2} \Big[w_c(u,m_l^2,m_M^2,\mu_{cut}^2) +
  w_d(u,m_l^2,m_M^2,\mu_{cut}^2)
\nonumber \\ & &
+
  w_f(u,m_l^2,m_M^2,\mu_{cut}^2)
-
  w_c(-u,m_l^2,m_M^2,\mu_{cut}^2)
\nonumber \\ & &
-
  w_d(-u,m_l^2,m_M^2,\mu_{cut}^2) -
  w_f(-u,m_l^2,m_M^2,\mu_{cut}^2) \Big]
\end{eqnarray}
Similarly the short distance correction to $R_{e/\mu}$ is
\be
  (\delta R_{e/\mu})_\subs{sd} = \frac{3}{2 f_M}
   \int_{-1}^{1} du \, \phi_M(u) r_{e/\mu}(u)
\ee
with
\be
  r_{e/\mu}(u) := 2 \left[\bar{w}(u,m_e^2,
   m_\pi^2,\mu_{cut}^2) - \bar{w}(u,m_\mu^2,m_\pi^2,\mu_{cut}^2)\right]
\ee
The relevant weight functions are given by
\begin{eqnarray}
   \lefteqn{w_c(u,m_l^2,m_M^2,\mu_{cut}^2)  =}
\nonumber \\
 & &   \frac{\alpha}{8\pi}
   \Bigg\{1 - 2 B_0 - 2 B_1 - 4 m_l^2 [B'_1 - B'_0]\Bigg\}
\nonumber \\ \nonumber \\
\lefteqn{
   w_d(u,m_l^2,m_M^2,\mu_{cut}^2) = }
\nonumber \\ & &
   \frac{\alpha}{6 \pi}
   \Bigg\{16 C_{00}^d + (1+u)^2 m_M^2 C_{11}^d + 4 m_l^2 C_{22}^d
   + 2 (1+u)(m_M^2 +  m_l^2) C_{12}^d
\nonumber \\ & & \qquad {}
   + [(1+u)^2 m_M^2 + (1+u) m_l^2] C_1^d
   + [(1+u)m_M^2 + (3+u)m_l^2] C_2^d
\nonumber \\ & & \qquad {}
   + (1+u) m_l^2 C_0^d - 2\Bigg\}
\nonumber \\ \nonumber\\
\lefteqn{
   w_f(u,m_l^2,m_M^2,\mu_{cut}^2)  =  }
\nonumber \\ & &
\frac{\alpha}{12 \pi}
   \Bigg\{4 C_{00}^f + (1-u)^2 m_M^2 C_{11}^f + 2(m_M^2 + m_l^2) C_{22}^f
   + (1-u)(3 m_M^2+m_l^2) C_{12}^f
\nonumber \\ & & \qquad {}
   + [(1-u)^2 m_M^2+(1-u)m_l^2] C_1^f
   + [(1-u) m_M^2+ (3-u) m_l^2] C_2^f
\nonumber \\ & & \qquad {}
   + (1-u) m_l^2 C_0^f \Bigg\}
\end{eqnarray}
where
\begin{eqnarray}
   B_{\dots}^{(\prime)} & = & B_{\dots}^{(\prime)}(m_l^2,\mu_{cut},m_l)
\nonumber \\[1ex]
   C_{\dots}^d & = & C_{\dots}\left(\frac{(1+u)^2}{4} m_M^2,
   \frac{1-u}{2} m_l^2 -
   \frac{1-u^2}{4} m_M^2, m_l^2,\mu_{cut},0,m_l\right)
\nonumber \\[1ex]
   C_{\dots}^f &= &C_{\dots}\left(\frac{(1-u)^2}{4} m_M^2,
   \frac{1+u}{2} m_l^2 -
   \frac{1-u^2}{4} m_M^2, m_l^2,\mu_{cut},0,m_l\right)
\end{eqnarray}

\begin{figure}
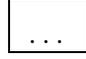

\caption{The function $r_{e/\mu}(u)$ for $\mu_{cut} = 1 \unit{GeV}$
(dashed), $\mu_{cut} = 1.5 \unit{GeV}$ (dotted)  and $\mu_{cut} =  2
\unit{GeV}$ (solid)}
\label{figwave3}
\begin{center}
\fbox{\str \ \dots \ }
  \end{center}
\end{figure}

In Figs.~\ref{figwave1} -- \ref{figwave3} we have plotted the symmetric
functions
$r_{\tau/\pi}(u)$, $r_{\tau/K}(u)$ and $r_{e/\mu}(u)$
for three different values of $\mu_{cut}$. It turns out that these
functions change only very little while varying $u$
 from $u=0$, where the distribution function $\phi_\pi$ is
presumably peaked, to $u=1$.
Therefore we can approximate the weight
functions by their values at $u=0$:
\begin{eqnarray}
   \frac{\delta \mcal_i}{\Gamma_0}(m_l^2, m_M^2, \mu_{cut}^2)
   & \approx & \frac{3}{2 f_M}
   w_i(0,m_l^2,m_M^2,\mu_{cut}^2) \int_{-1}^{+1} du \,
   \phi_M(u)
\nonumber \\ \nonumber \\
   & = & w_i(0, m_l^2, m_M^2,\mu_{cut}^2)
\end{eqnarray}
and then
\begin{eqnarray}
   \label{eqnfullsd}
  (\delta R_{\tau/M})_\subs{sd} &  = & \frac{3}{2 f_M}
   r_{\tau/M}(0) \int_{-1}^{1} du \, \phi_M(u)
\nonumber \\ \nonumber \\ & = &
   r_{\tau/M}(0)
\end{eqnarray}
It is clear from Figs.~\ref{figwave1} -- \ref{figwave3}, that this
approximation induces an uncertainty which is well below ${}_{\D -
0.00\%}^{\D +
0.02 \%}$ for  $R_{\tau/\pi}$, well below ${}_{\D-0.02\%}^{\D +0.00\%}$ for
$R_{\tau/K}$ and well below ${}_{\D -0.000\%}^{\D +0.003\%}$ for
$R_{e/\mu}$.
Note that we can make these statements without any specific
assumptions on the form of the
distribution functions $\phi_\pi$ and $\phi_K$.

\begin{figure}
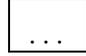

\caption{Short distance correction $(\delta R_{\tau/\pi})_{sd}$:
complete result according to \protect\eqn{eqnfullsd} (dotted) and
estimate based on the leading logarithms according to
\protect\eqn{eqnsdlog} (solid)}
\label{figsdlog}
\begin{center}
\fbox{\str \ \dots \ }
  \end{center}
\end{figure}

In \fig{figsdlog} we compare $(\delta R_{\tau/\pi})_{sd}$ as a function of
$\mu_{cut}$ as given by \eqn{eqnfullsd} with the
estimate from the leading logarithms
\be \label{eqnsdlog}
   (\delta R_{\tau/\pi})_{sd} \approx
   \frac{2 \alpha}{\pi} \frac{m_\tau^2}{m_\tau^2 - \mu_{cut}^2}
   \ln \frac{\mu_{cut}}{m_\tau}
\ee
We find that \eqn{eqnsdlog} gives an excellent approximation to
\eqn{eqnfullsd} and that the non-leading contributions to $(\delta
R_{\tau/\pi})_{sd}$ are very small.
%
\section{Hadronic Structure Dependent Loops}
\label{sec38}
\begin{figure}
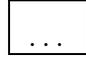

   \caption{Diagrams for hadronic structure dependent corrections}
   \label{figc7}
\begin{center}
\fbox{\str \ \dots \ }
  \end{center}
\end{figure}

%
For the real photon emission we have calculated the hadronic structure
dependent effects in \cite{Dec93}, cf.\ \sect{secrad}.
In the long distance virtual corrections
there are two different effects associated with hadronic
structure.
On the one hand, the photon emitted in the radiative decay by
hadronic structure dependent radiation (SD) as a real photon
could be reabsorbed either by the lepton or by the
pion (kaon), see \fig{figc7}. If in the radiative decay ($k^2 = 0$)
there is a
 SD amplitude proportional to $F_V$ and $F_A$ , the corresponding
hadronic ``structure dependent loops'' (SDL) must also be there for
sufficiently small $k^2$ (long distance).
Actually
for the last two diagrams, where the photon couples to the meson, the
respective corrections to tau and to  meson decay will be identical and
therefore they  will cancel in $R_{\tau/\pi}$.
So for simplicity we will not consider them.

\begin{figure}
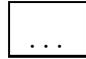

\caption{Vector meson dominance of coupling of the photon to the pion}
\label{figc8}
\begin{center}
\fbox{\str \ \dots \ }
\end{center}
\end{figure}

\begin{figure}
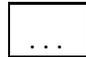

   \caption{Vector meson dominance of the photon coupling in the
   hadronic structure dependent diagrams}
   \label{figc9}
\begin{center}
\fbox{\str \ \dots \ }
  \end{center}
\end{figure}

On the other hand, according to the notion of vector meson dominance,
the photon does not couple directly to the pion (kaon)  but rather through a
$\rho$ intermediate state, see \fig{figc8}. Similarly in the diagrams
proportional to $F_V$ and $F_A$ the photon might couple through vector
meson dominance as indicated in \fig{figc9}.
In the remaining of this article these modifications of the photon
couplings
will {\em only} be called ``vector meson dominance'' of the respective
diagrams, either of the point meson loops (PML) or of the hadronic
structure dependent loops (SDL). The name ``hadronic structure
dependent'' will be used only for amplitudes such as $\delta
\mcal_6$ and $\delta \mcal_7$ which are proportional to the form
factors $F_V$ and $F_A$. This is of course just a naming convention;
the vector meson dominance of the photon coupling
is also an effect of hadronic structure.

We will discuss the modifications due to vector meson dominance of the
photon couplings later
and start with the calculation of $\delta
\mcal_6$ and $\delta \mcal_7$ using the generalized
photon propagator with  mass $m^2$
\be
   \frac{1}{k^2 - \lambda^2} \to \frac{1}{k^2 - m^2}
\ee
and no vector meson dominance in the coupling of the photon to the
vector meson and the pion (see Figs.~\ref{figc7} (a) and (b)).

The result for  $\delta\mcal_6$ is
\be
  \frac{\delta \mcal_6}{\mcal_0}(m_l^2, m_M^2, M_V^2, m^2)
   = \frac{\alpha}{2 \pi} \frac{F_V^M(0) M_V^2}
  {\sqrt{2} m_M f_M} \Big\{ 3 C_{00} + \frac{1}{2}(m_l^2 - m_M^2)
  (C_{22} + C_{12}) \Big\}
\ee
where
\be
   C_{ij} = C_{ij}(m_M^2, 0, m_l^2, m, M_V, m_l)
\ee
$M_V$ denotes a  -- possibly complex --- vector meson mass
\be
   M_V^2 = m_V^2 -i m_V \Gamma_V
\ee
A comment on the treatment of the meson propagators for the
$\rho$, $\rho'$, $a_1$, \dots particles is in order. In
the calculation of the radiative tau decay we used  sophisticated
Breit-Wigner resonance factors with energy dependent widths.
However, the difference between fixed and energy
dependent widths is very small after integration over the
spectrum. In the virtual corrections one has to integrate over all
possible loop momenta anyway. For the virtual corrections, we will
therefore only use
Breit-Wigner resonance factors with fixed widths
\be
   \BW_V(k^2) =
   \frac{M_V^2}{M_V^2 - k^2} = \frac{m_V^2 - i m_V \Gamma_V}
   {m_V^2 - k^2 - i m_V \Gamma_V}
\ee
or even with real masses $M_V^2 = m_V^2$.

Similarly the amplitude $\delta \mcal_7$ is equal to
\begin{eqnarray}
  \frac{\delta \mcal_7}{\mcal_0}(m_l^2, m_M^2, M_A^2, m^2)
   & = &\frac{\alpha}{4 \pi} \frac{F_A^{M}(0) M_A^2}
   {\sqrt{2} m_M f_M} \Big\{6 C_{00} + 3 m_M^2 C_{11} +
   (m_l^2 + 2 m_M^2) C_{22}
\nonumber \\ & &
 \quad {} + (2 m_l^2 + 4 m_M^2) C_{12}
   + (m_l^2 - m_M^2) C_{1} - 1 \Big\}
\end{eqnarray}
with
\be
  C_{\dots} = C_{\dots} ( m_M^2, 0 , m_l^2, m, M_A, m_l)
\ee
In the case of the meson being the pion,
our standard parameterization of the form factor $F^{(\pi)}_V(t)$ in the
radiative decay
was a tripole dominated by resonances $\rho$, $\rho'$ and $\rho''$ with
relative strengths $\sigma$ and $\rho$ (see \sect{secrad}). The
same parameterization will be used here for the structure dependent
loop $\delta \mcal_6$ , resulting in
\begin{eqnarray}
   \frac{\delta \mcal_V}{\mcal_0}(m_l^2, m_\pi^2, m^2)
    & = & \frac{1}{1+\sigma+\rho} \Big\{
    \frac{\delta \mcal_6}{\mcal_0}(m_l^2, m_\pi^2,M_\rho^2,m^2)
\nonumber \\ & & \qquad {}
\nonumber \\ & & \qquad {}
   + \sigma\,\,  \frac{\delta \mcal_6}
   {\mcal_0}(m_l^2, m_\pi^2,M_{\rho'}^2,m^2)
\nonumber \\ & & \qquad {}
\nonumber \\ & & \qquad {}
   + \rho\, \, \frac{\delta \mcal_6}
   {\mcal_0}(m_l^2, m_\pi^2,M_{\rho''}^2,m^2)
   \Big\}
\nonumber \\ & & \qquad {}
\nonumber \\
   \frac{\delta \mcal_A}{\mcal_0}(m_l^2, m_\pi^2, m^2)
   & = &
   \frac{\delta \mcal_7}{\mcal_0}(m_l^2, m_K^2,M_{a_1}^2,m^2)
\end{eqnarray}

and for the kaon we define
\begin{eqnarray}
   \frac{\delta \mcal_V}{\mcal_0}(m_l^2, m_K^2, m^2)
    & = & \frac{1}{1+\sigma_K+\rho_K} \Big\{
    \frac{\delta \mcal_6}{\mcal_0}(m_l^2, m_K^2,M_\kst^2,m^2)
\nonumber \\ & & \qquad {}
\nonumber \\ & & \qquad {}
   + \sigma_K\,\,  \frac{\delta \mcal_6}
   {\mcal_0}(m_l^2, m_K^2,M_{\kst'}^2,m^2)
\nonumber \\ & & \qquad {}
\nonumber \\ & & \qquad {}
   + \rho_K \, \, \frac{\delta \mcal_6}
   {\mcal_0}(m_l^2, m_K^2,M_{\kst''}^2,m^2)
   \Big\}
\nonumber \\
   \frac{\delta \mcal_A}{\mcal_0}(m_l^2, m_K^2, m^2)
   & = &
    \frac{\delta \mcal_7}{\mcal_0}(m_l^2, m_K^2,M_{K_1}^2,m^2)
\end{eqnarray}

Here again we use the convention that capital letters for masses
indicate possibly complex masses,
\be
   M_{a_1}^2 = m_{a_1}^2 - i m_{a_1} \Gamma_{a_1} \qquad
   M_{K_1}^2 = m_{K_1}^2 - i m_{K_1} \Gamma_{K_1}
\ee
and so on.
In terms of the rate the corrections are then  given by
\be
  \frac{\delta \Gamma_\subs{SDL}}{\Gamma_0} (m_l^2, m_M^2, m^2)
   = 2 \Re \left\{
  \frac{\delta \mcal_V}{\mcal_0} (m_l^2, m_M^2, m^2)
  + \frac{\delta \mcal_A}{\mcal_0} (m_l^2, m_M^2, m^2)
  \right\}
\ee

\begin{figure}
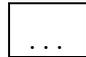

\caption{Vector meson dominance in the seagull coupling}
\label{figc8b}
\begin{center}
\fbox{\str \ \dots \ }%
\end{center}
\end{figure}

Now we will discuss diagram by diagram whether the coupling to
the photon should be modified by vector meson dominance.
Consider the effective point meson graphs in \fig{figc4}. In $\delta
\mcal_1$, the coupling $\gamma \pi\pi$ is determined by the
electromagnetic form factor of the pion, which is well know to be dominated
by the $\rho$ vector meson (see \fig{figc8}). The diagrams $\delta
\mcal_3$ and $\delta \mcal_4$ cancel in the ratio $R_{\tau/\pi}$, so we will
not include vector meson dominance here.
In diagram $\delta \mcal_5$ the photon
couples to the tau only, so the only diagram which remains to be considered
apart of $\delta \mcal_1$
is the diagram $\delta \mcal_2$ with the seagull coupling of the
photon to the weak interaction vertex.
Here it not clear at all whether this graph should be multiplied by a
rho Breit Wigner (compare \fig{figc8b}). However, care must be
taken to insure gauge invariance. The individual diagrams are not
gauge invariant, but their sum is. So the modification of $\delta
\mcal_2$ must be made in such a way that the sum of the diagrams is
gauge invariant.
This imposition of gauge invariance determines
that the correct modification of the diagram $\delta
\mcal_2$ is  given neither by the multiplication with 1 (ie.\ no
vector meson dominance, VMD) nor by the multiplication with $BW_\rho$
(complete VMD), but by the multiplication with
\be
   2 \BW_\rho(k^2) - 1
\ee
We will call this below the ``seagull type VMD''.

Next consider the corrections $\delta \mcal_6$ and $\delta \mcal_7$.
In these diagrams the coupling of the photon could be dominated by the
$\omega$ and the $\rho$ mesons, respectively (see \fig{figc9}).
We do not have any experimental information on whether or
not these couplings are actually dominated by vector mesons.
 (An experimental test could be made by
measuring the $e^+ e^-$ invariant mass spectra in the decays $\rho \to
\pi e^+ e^-$ and $a_1 \to \pi e^+ e^-$, respectively.) By
extrapolation from the experience with other hadronic couplings of the
photon one could expect vector meson dominance here as well, but
in order to be unprejudiced we
will below consider both possibilities, complete VMD and  no VMD.

Now according to \sect{sec32}, in the long distance diagrams a
regularized photon propagator
\be
   \frac{1}{k^2 - \lambda^2} \to \frac{1}{k^2-\lambda^2}
   \frac{\mu_{cut}^2}{\mu_{cut}^2 - k^2}
\ee
is to be used. If the photon additionally couples via vector meson
dominance with a vector resonance mass $M_R$, the simple photon
propagator has to be replaced by
\be
   \frac{1}{k^2 - \lambda^2} \to \frac{1}{k^2-\lambda^2}
   \frac{\mu_{cut}^2}{\mu_{cut}^2 - k^2} \frac{M_R^2}{M_R^2 - k^2}
\ee

Assume a long distance diagram has been calculated with the photon
propagator
\be \label{eqn388}
   \frac{1}{k^2 - m^2}
\ee
and without vector meson dominance coupling of the photon, with the
result $G(m^2)$.
Then the following replacements have to be made in order to get the
correct answer:
\begin{itemize}
\item long distance diagram without VMD
\be
   G(m^2) \to G(\lambda^2) - G(\mu_{cut}^2)
\ee
\item long distance diagram with usual VMD
\be
   G(m^2) \to G(\lambda^2) - G(\mu_{cut}^2)
   + \frac{\mu_{cut}^2}{\mu_{cut} - M_R^2} \Big[ G(\mu_{cut}^2)
   - G(M_R^2) \Big]
\ee
\item long distance diagram with seagull type VMD
\be \label{eqn391}
   G(m^2) \to G(\lambda^2) - G(\mu_{cut}^2)
   + \frac{2 \mu_{cut}^2}{\mu_{cut} - M_R^2} \Big[ G(\mu_{cut}^2)
   - G(M_R^2) \Big]
\ee
\end{itemize}
%
\section{Complete Radiative Correction}
%
We write the complete radiative correction as
\begin{eqnarray}
   \delta R_{\tau/M} & = &
   (\delta R_{\tau/M})_\subs{CPM}
   + (\delta R_{\tau/M})_\subs{VMD(PML)}
   + (\delta R_{\tau/M})_\subs{HSD}
\nonumber \\ & & {}
\nonumber \\ & & {}
   + (\delta R_{\tau/M})_\subs{VMD(HSD)}
   + (\delta R_{\tau/M})_\subs{sd}
\end{eqnarray}
using the naming conventions
\begin{itemize}
\item
  ``CPM'' = cut point meson,
   ie.\ the long distance correction due to an effective point meson,
   including real and virtual photonic correction, where the loops
   have been calculated with the regulated photon propagator with
   the cut-off scale $\mu_{cut}^2$
\item
  ``VMD(PML)'' = vector meson dominance of the point meson loops,
  ie.\ the correction of the CPM result due to the vector
  meson dominance of the meson electromagnetic form factor,
\item
   ``HSD'' = hadronic structure dependent, ie.\ the correction due
   to the diagrams proportional to $F_V$ and $F_A$, including
   real and virtual corrections,
\item
   ``VMD(SDL)'' = vector meson dominance of the structure dependent
   loops, ie.\ the modification of the HSD result due to vector
   meson dominance coupling of the photon in these diagrams, and finally
\item
  ``sd'', the short distance correction.
\end{itemize}
For the integrated rates of the real photon emission we divide into pure
internal bremsstrahlung (IB) and the rest (SD + INT), viz.\ the sum of
pure structure dependent radiation and of the interference between the
internal bremsstrahlung and the structure dependent radiation.

And so
\begin{eqnarray}
   (\delta R_{\tau/M})_\subs{CPM} & = &
   \frac{\Gamma_\subs{IB}(\tau\to M\nu_\tau\gamma)}
   {\Gamma_0(\tau\to M \nu_\tau)}
   - \frac{\Gamma_\subs{IB}(M \to \mu \nu_\mu \gamma)}
   {\Gamma_0(M \to \mu \nu_\mu)}
\nonumber \\ & &
\nonumber \\ & &
   + \frac{\delta \Gamma_\subs{PML}}{\Gamma_0}(m_\tau^2, m_M^2,\lambda^2)
   - \frac{\delta \Gamma_\subs{PML}}{\Gamma_0}(m_\tau^2, m_M^2, \mu_{cut}^2)
\nonumber \\ & &
\nonumber \\ & &
   - \frac{\delta \Gamma_\subs{PML}}{\Gamma_0}(m_\mu^2, m_M^2,\lambda^2)
   + \frac{\delta \Gamma_\subs{PML}}{\Gamma_0}(m_\mu^2, m_M^2,\mu_{cut}^2)
\end{eqnarray}
and
\begin{eqnarray}
   (\delta R_{\tau/M})_\subs{HSD} & = &
   \frac{\Gamma_\subs{SD+INT}(\tau\to M\nu_\tau\gamma)}
   {\Gamma_0(\tau\to M \nu_\tau)}
   - \frac{\Gamma_\subs{SD+INT}(M \to \mu \nu_\mu \gamma)}
   {\Gamma_0(M \to \mu\nu_\mu)}
\nonumber \\ & &
\nonumber \\ & &
   + \frac{\delta \Gamma_\subs{SDL}}{\Gamma_0}(m_\tau^2, m_M^2,\lambda^2)
   - \frac{\delta \Gamma_\subs{SDL}}{\Gamma_0}(m_\tau^2, m_M^2, \mu_{cut}^2)
\nonumber \\ & &
\nonumber \\ & &
   - \frac{\delta \Gamma_\subs{SDL}}{\Gamma_0}(m_\mu^2, m_M^2,\lambda^2)
   + \frac{\delta \Gamma_\subs{SDL}}{\Gamma_0}(m_\mu^2, m_M^2,\mu_{cut}^2)
\end{eqnarray}

For the point meson loops the vector meson dominance is taken into
account by
\be
  (R_{\tau/M})_\subs{VMD(PML)} =
  \frac{\Gamma_\subs{VMD(PML)}}{\Gamma_0}(m_\tau^2,m_M^2)
 -   \frac{\Gamma_\subs{VMD(PML)}}{\Gamma_0}(m_\mu^2,m_M^2)
\ee
where according to the last section
\begin{eqnarray}
\lefteqn{  \frac{\delta \Gamma_\subs{VMD(PML)}}{\Gamma_0}
(m_l^2, m_M^2)
= \frac{2}{1 + \sigma_1
  + \rho_1} \Re\Bigg\{ }
\nonumber \\ & &
  \frac{\mu_{cut}^2}{\mu_{cut}^2 - M_\rho^2}
  \Bigg[ \frac{\delta\mcal_1}{\mcal_0}(m_l^2,m_M^2,\mu_{cut}^2)
  + 2 \frac{\delta\mcal_2}{\mcal_0}(m_l^2,m_M^2,\mu_{cut}^2)
\nonumber \\ & & \qquad  \qquad \quad
- \frac{\delta\mcal_1}{\mcal_0}(m_l^2,m_M^2,M_\rho^2)
  - 2 \frac{\delta\mcal_2}{\mcal_0}(m_l^2,m_M^2,M_\rho^2)
  \Bigg]
\nonumber \\ & &
  + \sigma_1 \frac{\mu_{cut}^2}{\mu_{cut}^2 - M_{\rho'}^2}
  \Bigg[ \frac{\delta\mcal_1}{\mcal_0}(m_l^2,m_M^2,\mu_{cut}^2)
  + 2 \frac{\delta\mcal_2}{\mcal_0}(m_l^2,m_M^2,\mu_{cut}^2)
\nonumber \\ & & \qquad  \qquad \quad
- \frac{\delta\mcal_1}{\mcal_0}(m_l^2,m_M^2,M_{\rho'}^2)
  - 2 \frac{\delta\mcal_2}{\mcal_0}(m_l^2,m_M^2,M_{\rho'}^2)
  \Bigg]
\nonumber \\ & &
  + \rho_1 \frac{\mu_{cut}^2}{\mu_{cut}^2 - M_{\rho''}^2}
  \Bigg[ \frac{\delta\mcal_1}{\mcal_0}(m_l^2,m_M^2,\mu_{cut}^2)
  + 2 \frac{\delta\mcal_2}{\mcal_0}(m_l^2,m_M^2,\mu_{cut}^2)
\nonumber \\ & & \qquad  \qquad \quad
- \frac{\delta\mcal_1}{\mcal_0}(m_l^2,m_M^2,M_{\rho''}^2)
  - 2 \frac{\delta\mcal_2}{\mcal_0}(m_l^2,m_M^2,M_{\rho''}^2)
  \Bigg] \Bigg\}
\end{eqnarray}
where the parameters $\sigma_1$ and $\rho_1$ describe the relative
contribution of the $\rho'$ and the $\rho''$ in the
electromagnetic form factor of the meson
(compare
$\sigma$ and $\rho$ in the vector form factor $F_V$).
Note that for the case of the kaon, $m_M = m_K$ , this assumes
complete $U(3)$ flavour symmetry in the vector meson sector, $M_\rho =
M_\omega = M_\phi$. Otherwise the relative contributions of the
$\rho^0$, the $\omega$ and the $\Phi$ in the electromagnetic form
factor of the kaon would have to be considered.

The vector meson dominance in the ``hadronic structure dependent
loops'' (SDL) is implemented by
\be
  (R_{\tau/M})_\subs{VMD(SDL)} =
  \frac{\Gamma_\subs{VMD(SDL)}}{\Gamma_0}(m_\tau^2,m_M^2)
 -   \frac{\Gamma_\subs{VMD(SDL)}}{\Gamma_0}(m_\mu^2,m_M^2)
\ee
with
\begin{eqnarray}
\lefteqn{  \frac{\delta \Gamma_\subs{VMD(SDL)}}{\Gamma_0}(m_l^2, m_M^2)
  = \frac{2 f_2}{1 + \sigma_2 + \rho_2} \Re\Bigg\{ }
\nonumber \\ & &
  \frac{\mu_{cut}^2}{\mu_{cut}^2 - M_\omega^2}
  \Bigg[ \frac{\delta\mcal_V}{\mcal_0}(m_l^2,m_M^2,\mu_{cut}^2)
- \frac{\delta\mcal_V}{\mcal_0}(m_l^2,m_M^2,M_\omega^2)
  \Bigg]
\nonumber \\ & &
  + \sigma_2 \frac{\mu_{cut}^2}{\mu_{cut}^2 - M_{\omega'}^2}
  \Bigg[ \frac{\delta\mcal_V}{\mcal_0}(m_l^2,m_M^2,\mu_{cut}^2)
- \frac{\delta\mcal_V}{\mcal_0}(m_l^2,m_M^2,M_{\omega'}^2)
  \Bigg]
\nonumber \\ & &
  + \rho_2 \frac{\mu_{cut}^2}{\mu_{cut}^2 - M_{\omega''}^2}
  \Bigg[ \frac{\delta\mcal_V}{\mcal_0}(m_l^2,m_M^2,\mu_{cut}^2)
- \frac{\delta\mcal_V}{\mcal_0}(m_l^2,m_M^2,M_{\omega''}^2)
  \Bigg] \Bigg\}
\nonumber \\
  \lefteqn{+ \frac{2 f_3}{1 + \sigma_3 + \rho_3} \Re\Bigg\{ }
\nonumber \\ & &
  \frac{\mu_{cut}^2}{\mu_{cut}^2 - M_\rho^2}
  \Bigg[ \frac{\delta\mcal_A}{\mcal_0}(m_l^2,m_M^2,\mu_{cut}^2)
- \frac{\delta\mcal_A}{\mcal_0}(m_l^2,m_M^2,M_\rho^2)
  \Bigg]
\nonumber \\ & &
  + \sigma_3 \frac{\mu_{cut}^2}{\mu_{cut}^2 - M_{\rho'}^2}
  \Bigg[ \frac{\delta\mcal_A}{\mcal_0}(m_l^2,m_M^2,\mu_{cut}^2)
- \frac{\delta\mcal_A}{\mcal_0}(m_l^2,m_M^2,M_{\rho'}^2)
  \Bigg]
\nonumber \\ & &
  + \rho_3 \frac{\mu_{cut}^2}{\mu_{cut}^2 - M_{\rho''}^2}
  \Bigg[ \frac{\delta\mcal_A}{\mcal_0}(m_l^2,m_M^2,\mu_{cut}^2)
- \frac{\delta\mcal_A}{\mcal_0}(m_l^2,m_M^2,M_{\rho''}^2)
  \Bigg] \Bigg\}
\end{eqnarray}
where $f_2$ and $f_3$ are flags which are either one or zero,
determining whether or not the photon coupling to the vector
resonance and the pion or to the axial resonance and the pion are
dominated by $\omega$ and $\rho$ type resonances, respectively. The
parameters $\sigma_2$ and $\rho_2$ for the $\omega$ type ones and
$\sigma_3$ and $\rho_3$ for the $\rho$ type ones describe the relative
contributions of higher radial excitations.

{}From our results we can also calculate the correction $\delta R_{e/\mu}$
to the ratio of the electronic and the muonic decay modes of the pion:
\begin{eqnarray}
   \delta R_{e/\mu} & = &
   (\delta R_{e/\mu})_\subs{CPM}
   + (\delta R_{e/\mu})_\subs{VMD(PML)}
   + (\delta R_{e/\mu})_\subs{HSD}
\nonumber \\ & & {}
\nonumber \\ & & {}
   + (\delta R_{e/\mu})_\subs{VMD(HSD)}
   + (\delta R_{e/\mu})_\subs{sd}
\end{eqnarray}
where
$(\delta R_{e/\mu})_{CPM}$, \dots, $(\delta R_{e/\mu})_{sd}$ are defined
and calculated in a completely analogous way.
%
\section{Numerical Results}
%
We will now evaluate the formulae of the last section numerically and
so get our final results.
For the calculation of the standard loop integrals we have used the
numerical programs FF \cite{Old90} and AA/FF \cite{Aep93}.

Uncertainties of the final result come from two different sources: on
the one hand from uncertainties in the hadronics and on the other hand
from uncertainties in the matching of long and short distances.
We estimate their sizes by varying the various parameters involved,
ie.\ $F_A^\pi(0)$, relative contributions of higher radial
resonances and others for the hadronics and by varying $\mu_{cut}$ for
the matching.
We obtain a central value with an error
estimate in the following way. We discuss the various parameters
and find for each one of them a central value
and some reasonable range over which we will vary them. Taking all
parameters at their central values, we obtain the central value for
the total radiative correction. We then vary the various parameters
and determine  which choices give the smallest and  the largest
radiative correction, respectively.
Then by simultaneously taking those values which
result in the largest (smallest) correction, we obtain an upper
(lower) limit for the radiative correction.

\begin{table} 
\caption{Parameter sets for $R_{\tau/\pi}$}
\label{tabparataupi}
$$
  \begin{array}{c|ccc}
  \mbox{Parameter} & \mbox{I} & \mbox{II} & \mbox{III} \\
  \hline
  F_A(0) = & 0.0116 & 0.0100 & 0.0132 \\
  \Gamma_{a_1} \,[\mbox{MeV}] = & 400 & 600 & 250 \\
  \sigma = & 0.136 & 0.0584 & 0.000\\
  \rho   = & -0.051 & 0.0 & 0.000 \\
  \mbox{widths in real radiation: $F_V$} & \mbox{energy dependent} &
  \mbox{energy dependent}  & \mbox{fixed} \\
  \mbox{widths in real radiation: $F_A$} & \mbox{energy dependent} &
  \mbox{fixed}  & \mbox{energy dependent} \\
  \sigma_1= & -0.1 & -0.1 & 0 \\
  \rho_1= & -0.04 & -0.04 & 0 \\
  f_2 = f_3= & 1 & 0 & 1 \\
  \sigma_2 = \sigma_3= & 0 &  \mbox{---} & -0.1 \\
  \rho_2 = \rho_3 = & 0 &  \mbox{---} & -0.04\\
\end{array}
$$
\end{table}
\begin{table} 
\caption{Parameter sets for $R_{\tau/K}$}
\label{tabparatauka}
$$
  \begin{array}{c|ccc}
  \mbox{Parameter} & \mbox{I} & \mbox{II} & \mbox{III} \\
  \hline
  F_A^{(K)}(0) = & 0.0525 & 0.0425 & 0.0625 \\
  \Gamma_{K_1} \,[\mbox{MeV}] = & 90 & 110 & 70 \\
  \sigma_K = & 0.000 & 0.136 & 0.000\\
  \rho_K   = & 0.000 & -0.051 & 0.000 \\
%
  \mbox{widths in real radiation} & \mbox{fixed} & \mbox{fixed}
   & \mbox{fixed} \\
  \sigma_1= & 0 & -0.1 & 0 \\
  \rho_1= & 0 & -0.04 & 0 \\
  f_2 = f_3= & 1 & 0 & 1 \\
  \sigma_2 = \sigma_3= & 0 & \mbox{---} & -0.1\\
  \rho_2 = \rho_3 =& 0 &  \mbox{---} & -0.04\\
\end{array}
$$
\end{table}
\begin{table} 
\caption{Parameter sets for $R_{e/\mu}$}
\label{tabparaelmu}
$$
  \begin{array}{c|ccc}
  \mbox{Parameter} & \mbox{I} & \mbox{II} & \mbox{III} \\
  \hline
  F_A(0) = & 0.0116 & 0.0100 & 0.0132 \\
  \Gamma_{a_1} \,[\mbox{MeV}] = & 400 & 600 & 250 \\
  \sigma = & 0.136 & 0.0584 & 0.000\\
  \rho   = & -0.051 & 0.0 & 0.000 \\
  \mbox{widths in real radiation: $F_V$} & \mbox{energy dependent} &
  \mbox{energy dependent}  & \mbox{fixed} \\
  \mbox{widths in real radiation: $F_A$} & \mbox{energy dependent} &
  \mbox{fixed}  & \mbox{energy dependent} \\
  \sigma_1= & -0.1 & 0 & -0.1 \\
  \rho_1= & -0.04 & 0 & -0.04 \\
  f_2 = f_3= & 1 & 1 & 0 \\
  \sigma_2 = \sigma_3= & 0 & -0.1 & \mbox{---} \\
  \rho_2 = \rho_3 =& 0 & -0.04 & \mbox{---}  \\
\end{array}
$$
\end{table}
For the parameters of the hadronics, we obtain in this way three
parameter sets (I)--(III), we are given is the tables
\ref{tabparataupi}, \ref{tabparatauka} and \ref{tabparaelmu},
corresponding to the central values (I) and to the lower (II) and
upper limits (III)  on the total radiative correction.

For the scale $\mu_\subs{cut}$
the range $\mu_{cut} = 1 \dots 2 \unit{GeV}$ is reasonable and we will
use $
   \mu_{cut} = 1.5 \unit{GeV}
$ as an intermediate standard value.

If all parameters have their standard values we find the total
radiative correction
\be
   R_{\tau/\pi} = 0.16\%
\ee

Because of all the ambiguities of the non-perturbative strong
interactions in the $O(1\unit{GeV})$ regime, it is  interesting
to consider the result which is obtained if in the long
distance part only the point meson is used, without vector meson
dominance and without hadronic structure dependent loops.
This point pion contribution is fixed by the low energy theorems of
QCD and therefore free from any ambiguities.
However, in
this case the result depends very strongly on the value of the scale
$\mu_{cut}$. This is shown in \fig{figm1} where we show the
correction $\delta R_{\tau/\pi}$ in variation with the scale $\mu_{cut}$ both
for our complete calculation, using standard parameters, and for the
calculation where in the long distance part only the point meson is
taken into account. In the last case the dependence on $\mu_{cut}$ is
very strong. As has been discussed before, the pure point meson is
reliable only for $\mu_{cut}^2 \ll m_\rho^2$, but the short
distance correction only for $\mu_{cut}\, {}_\sim^>\, (1 \dots 2)
\unit{GeV}$. So here a very large range for $\mu_{cut}$ has to be
considered. For a small $\mu_{cut}$ the correction $\delta R_{\tau/\pi}$
could even become negative, and at $\mu_{cut} = 2 \unit{GeV}$ the
correction is about $\delta R_{\tau/\pi} = 0.7 \%$ and still rising
strongly.

This large dependence on $\mu_{cut}$ results from the incomplete
treatment of the long distance part and hints to the necessicity of
improving the model.
The inclusion of VMD and of the structure dependent loops decreases
the dependence on $\mu_{cut}$ very much, as can be seen from the other
curve in \fig{figm1}. Above $\mu_{cut} = 2 \unit{GeV}$ the curve
becomes almost completely flat, and the variation in the (relevant)
range
$\mu_{cut} = (1 \dots 2) \unit{GeV}$ is smaller than $\pm 0.05\%$.
%
\addtolength{\topmargin}{-2cm}
\begin{footnotesize}
\begin{table}
\caption{The different contributions adding up to the total radiative
correction $\delta R_{\tau/\pi}$ for $\mu_{cut} = 1.5 \unit{GeV}$}
\label{tabeinzeln}
\begin{center}
\begin{tabular}{|l| c| c|}
\hline \str
\parbox[c]{5cm}{\str effective point pion\\} &
$    (\delta R_{\tau/\pi })_\subs{PM}  $ & +1.05 \% \str \\ \hline \str
\parbox[c]{5cm}{cutting off the point pion loops at $\mu_{cut}$} &
 \parbox[c]{5cm}{$$ - \frac{\delta
   \Gamma_\subs{PML}}{\Gamma_0}(m_\tau^2, m_\pi ^2, \mu_{cut}^2) $$ $$
   + \frac{\delta \Gamma_\subs{PML}}{\Gamma_0}(m_\mu^2, m_\pi
^2,\mu_{cut}^2)$$}
 & $- 0.21 \%$ \str \\
\hline \str
\parbox[c]{5cm}{\str vector meson dominance of the pion electromagnetic
form factor\\ }  &
$ (R_{\tau/\pi })_\subs{VMD(PML)} $ &
$- 0.38 \%$
\str \\  \hline \str
\parbox[c]{5cm}{ Structure dependent  radiation and SD-IB
interference (real photon emission)\\} &
\parbox[c]{5cm}{ $$
   \frac{\Gamma_\subs{SD+INT}(\tau\to \pi \nu_\tau\gamma)}
   {\Gamma_0(\tau\to \pi  \nu_\tau)}    $$ $$
   - \frac{\Gamma_\subs{SD+INT}(\pi  \to \mu \nu_\mu \gamma)}
   {\Gamma_0(\pi  \to \mu\nu_\mu)}
$$} &
$ +0.05 \% $
\str \\  \hline \str
\parbox[c]{5cm}{\str hadronic structure dependent loops, cut off at
$\mu_{cut}$ \\} &
\parbox[c]{5cm}{ $$
     \frac{\delta\Gamma_\subs{SDL}}{\Gamma_0}(m_\tau^2,m_\pi^2,0)
$$ $$
 - \frac{\delta \Gamma_\subs{SDL}}{\Gamma_0}(m_\tau^2, m_\pi^2, \mu_{cut}^2)\\
$$ $$
   - \frac{\delta \Gamma_\subs{SDL}}{\Gamma_0}(m_\mu^2, m_\pi^2,0)\\
$$ $$
   + \frac{\delta \Gamma_\subs{SDL}}{\Gamma_0}(m_\mu^2, m_\pi^2,\mu_{cut}^2)\\
 $$} &
$ -0.24 \% $
\str \\  \hline \str
\parbox[c]{5cm}{\str vector meson dominance of photon couplings in the
hadronic structure dependent loops \\} &
\parbox[c]{5cm}{ $$ (\delta R_{\tau/\pi })_\subs{VMD(SDL)}
$$} &
$ + 0.13 \% $
\str \\  \hline \str
\parbox[c]{5cm}{\str short distance contribution for $k^2 > \mu_{cut}^2$\\} &
\parbox[c]{5cm}{ $$  (\delta R_{\tau/\pi })_\subs{sd} $$} &
$ -0.25 \% $
\str \\  \hline \str
\parbox[c]{5cm}{\str sum \\} &
\parbox[c]{5cm}{ $$    \delta R_{\tau/\pi }  $$} &
$ +0.16 \% $
\str \\ \hline
\end{tabular} \end{center}
\comment{
\str \\  \hline \str
\parbox[c]{5cm}{ \\} &
\parbox[c]{5cm}{ $$ $$} &
$  $
}
\end{table}
\end{footnotesize}
\addtolength{\topmargin}{2cm}
In \tab{tabeinzeln} we show the different contributions which add up
to the total radiative correction, using $\mu_{cut} = 1.5 \unit{GeV}$.
The individual contributions involve moderately large logarithms
such as $\ln (m_\tau/ m_\pi)$ or $\ln(m_\rho/m_\tau)$. However they
have  opposite signs, such that most of the
corrections cancel and only a very small total radiative
correction of $\delta R_{\tau/\pi} = + 0.16 \%$ is left.

\begin{figure}
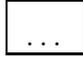

\caption{The different contributions which add up to the total
correction: The point pion correction, supplemented with vector meson
dominance in the photon coupling (dashed), the hadronic structure
dependent correction with vector meson dominance in the photon
coupling (dash-dotted), short distance correction (dotted) and their
sum, the total correction (solid)}
\label{figm2}
\begin{center}
\fbox{\str \ \dots \ }
  \end{center}
\end{figure}

In \fig{figm2} we show  the long and short distance corrections in
variation with $\mu_{cut}$. We display
separately the correction due to the effective point meson with
vector meson dominance coupling in the photon coupling
\be
   (\delta R_{\tau/M})_\subs{CPM}
   + (\delta R_{\tau/M})_\subs{VMD(PML)}
\ee
the hadronic structure dependent correction with vector meson dominance
\be
  (\delta R_{\tau/M})_\subs{HSD}
   + (\delta R_{\tau/M})_\subs{VMD(HSD)}
\ee
and the short distance correction
\be
    (\delta R_{\tau/M})_\subs{sd}
\ee
It can be seen clearly that the long and the short distance
corrections vary in the opposite way with $\mu_{cut}$, such
that the dependence of the total radiative correction on $\mu_{cut}$
is significantly smaller than that of the individual corrections
separately. This is a sensible result. If our long and short distance
corrections would exactly describe the real world within an
overlapping region of $\mu_{cut}$, the sum of the two would be
independent of $\mu_{cut}$ within this overlap region. The small
remaining dependence of our final result on $\mu_{cut}$ within the
range $\mu_{cut} = 1.0 \dots 2.0 \unit{GeV}$ indicates that our model
is not unreasonable.

In \fig{figtaupi} we show $\delta R_{\tau/\pi}$
in variation with $\mu_{cut}$ for
three different choices for the parameters of the resonance physics,
viz.\ for the standard set (I) and for the set (II) which gives the
smallest correction and for the set (III), which gives the largest
correction (see \tab{tabparataupi}).
All the three curves have been obtained using real vector meson
masses
(narrow width approximation) in the virtual corrections, but we have
compared with the results obtained with complex vector meson masses,
and in all cases the difference is extremely small and completely
negligible.
While the two curves for (I) and
(II) are close together, the curve for (III) lies significantly
higher. The large difference is due to the question whether or not the
photon in the hadronic structure dependent loops couples to the mesons
via vector meson dominance, all other parameter variations have a much
smaller influence.

As our final result on $\delta R_{\tau/\pi}$ we get from the parameter
sets (I) --- (III) and from varying $\mu_{cut} = 1.0 \dots 2.0
\unit{GeV}$:
\be
 \fbox{$ \D \delta R_{\tau/\pi} = \left(0.16_{\D -0.14}^{\D +0.09}
  \right) \%$}
\ee
In \fig{figtauka} we show $\delta R_{\tau/K}$ in variation with
$\mu_{cut}$ for the three parameter sets (I)--(III) which are defined
in \tab{tabparatauka}.
Our final result for the correction to $R_{\tau/K}$ is
\be
 \fbox{$ \D \delta R_{\tau/K} = \left(0.90_{\D -0.26}^{\D +0.17}\right) \%$}
\ee
where the central value is for the parameter set (I) and with
$\mu_{cut} = 1.5 \unit{GeV}$, and the lower and
upper limits are from the sets (II) and (III) and $\mu_{cut} =
1.0\unit{GeV},\, 2.0 \unit{GeV}$, respectively.

For the normalized branching ratios this results in
\begin{eqnarray}
   \frac{\br_\pi}{\br_e} & = & 0.6129 \pm \overbrace{0.0007}^\subs{expt}
    \overbrace{{}_{\D -0.0009}^{\D +0.0005}}^\subs{theo}
   = 0.6129_{\D -0.0011}^{\D +0.0009}
\nonumber \\
   \frac{\br_K}{\br_e} & = & 0.0406 \pm \overbrace{0.0002}^\subs{expt}
   \overbrace{{}_{\D -0.0001}^{\D +0.0000}}^\subs{theo}
   = 0.0406 \pm 0.0002
\nonumber \\
 \Rightarrow \frac{\br_{\pi + K}}{\br_e} & = & 0.6535 \pm
  \overbrace{0.0007}^\subs{expt} \overbrace{{}_{\D -0.0010}^{\D
   +0.0005}}^{theo}
  = 0.6535_{\D -0.0012}^{\D +0.0009}
\end{eqnarray}
The first errors given (called ``experimental'')
are due to the uncertainties in the lifetimes of
the mesons and in the tau mass $m_\tau$,
and the second errors (called ``theoretical'')
are due to the uncertainties in the
radiative correction. This final result deviates from the
experimental result \cite{Cad93} given in the introduction
\be
   \left(\frac{\br_{\pi + K}}{\br_e}\right)_{exp} =
   \frac{(11.99 \pm 0.25) \%}{(17.76 \pm 0.15) \%} = 0.675 \pm 0.015
\ee
by $1.4$ standard deviations.
The agreement between theory and
experiment is not significantly enhanced by the inclusion of the
$O(\alpha)$ corrections to the decay rate for $\tau\to\pi\nu_\tau$.
Thus if the standard model is
correct, either the current experimental number for the branching
ratio $\br_{\pi+K}$ is slightly too large or the one for $\br_{e}$
slightly too small, or both.

Some comments on the reliability of our matching procedure are in order.
First it is important to note that indeed the dependence of the long
and short distance corrections separately on $\mu_{cut}$ is
considerably larger than that of their sum, as has been discussed
above in connection with Fig.~\ref{figm2}. Second we have performed
the matching in a certain way be splitting the photon propagator into
a long and a short distance part according to \eqn{eqn36}. This
corresponds to a soft transition from long to short distances. Another
way to perform the matching would be a sharp transition from long
to short distances at $\mu_{cut}$, by integrating the
long distances from $k_E^2 = 0 \dots \mu_{cut}^2$ and the short
distances from $k_E^2 = \mu_{cut}^2 \dots m_Z^2$. In principle
the results obtained with this method might differ somewhat from our
results.
However, the uncertainty of
our final result for $\delta R_{\tau/\pi}$ is dominanted by the
hadronic uncertainties and not by the matching uncertainties. Varying
the parameters of the hadronics and the matching scale $\mu_{cut}$
separately, we obtain
\be
    \delta R_{\tau/\pi} = \left(0.16\,
    \overbrace{{}_{\D -0.05}^{\D +0.02}}^{\subs{matching}} \,
    \overbrace{{}_{\D -0.13}^{\D +0.07}}^\subs{hadronics}
    \right) \%
\ee
which clearly displays the dominance of the hadronic uncertainties.
Therefore we think that the precise procedure for performing the
matching is not essential. Still it might be an interesting task to
repeat the calculation with a matching based on a sharp transition at
$\mu_{cut}$.

\begin{figure}
\caption{The total radiative correction $R_{e/\mu}$
for different choices for the
parameters in the structure dependent correction: Standard choice (I)
(solid), choice (II) (dashed) and choice (III) (dotted)}
\label{figelmu}
\begin{center}
\fbox{\str \ \dots \ }
  \end{center}
\end{figure}
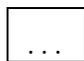
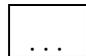
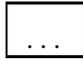
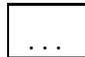

Now we will use our results to predict for the pion decay the ratio
$R_{e/\mu}$ of  the electronic and muonic modes.
In \fig{figelmu} we show $\delta R_{e/\mu}$ in variation with
$\mu_{cut}$ for  three parameter sets (I)---(III) defined in
\tab{tabparaelmu}.
We find for the total radiative correction
\be \label{eqn3121}
 \delta R_{e/\mu} = \left(-3.79 \pm 0.01\right) \%
\ee
But note that
this error, which is due to the uncertainties in $\mu_{cut}$ and in
the hadronic resonance physics, is smaller than the one which is to be
expected because of
the neglect higher order corrections of $O(\alpha^2)$, as we will
explain below.

The value in \eqn{eqn3121} results in
\begin{eqnarray}
   R_{e/\mu} & = & R_{e/\mu}^{(0)} \Big( 1 + \delta R_{e/\mu} \Big)
\nonumber \\
  & = & \frac{m_e^2}{m_\mu^2} \left(\frac{m_\pi^2 - m_e^2}
  {m_\pi^2 - m_\mu^2} \right)^2 \Big( 1 + \delta R_{e/\mu} \Big)
\nonumber \\
 & = & 1.2835 \cdot 10^{-4} \Big(1 + \left(-3.79 \pm 0.01
  \right) \%\Big)
\nonumber \\
 & = & (1.2349 \pm 0.0001) \cdot 10^{-4}
\end{eqnarray}
where the error given includes only the uncertainties resulting from
the hadronic structure dependent corrections and the matching scale
$\mu_{cut}$ and misses the comparable uncertainty resulting from the
neglect of the $O(\alpha)$ corrections.

This result  is to be compared with the recent result given by
Marciano and Sirlin \cite{Mar93}:
\be
   R_{e/\mu} = (1.2352 \pm 0.0005) \cdot 10^{-4}
\ee
which is equivalent to
\be
   \delta R_{e/\mu} = (- 3.76 \pm 0.04) \%
\ee
Their calculation is rather similar to ours but differs in some
details. In the short distance corrections they use the leading
logarithms only, supplemented by leading QCD corrections and
furthermore by summing up the leading logarithms to all orders in
$\alpha$ via the renormalization group. But since the
short distance corrections cancel almost completely in $R_{e/\mu}$,
these
differences are not important. For the long distance corrections they
also consider both the effective point pion field and hadronic
structure dependent corrections. However, for the latter they do not
use a specific model but rather use a theorem given by Terent'ev
\cite{Ter73} on the leading lepton mass dependent hadronic structure
effects of the order $\frac{\alpha}{\pi}\frac{ m_l^2}{m_\rho^2}  \ln
\frac{m_\rho^2}{m_l^2}$. The possible size of the  remaining hadronic
structure dependent corrections, which they do not calculate,
represents the main source of the uncertainty in their $R_{e/\mu}$ as
quoted above.
The authors also consider the effect of corrections of higher order in
$\alpha$. Since the correction $3 (\alpha/\pi) \ln (m_\mu/m_e) \approx
- 3.7 \%$, which results from the lepton mass singularities, dominates
the $O(\alpha)$ correction,
one expects its higher order counterparts to similarly dominate their
respective orders. Summing all such logarithms via the renormalization
group gives the enhancement factor
\be \label{eqn3125}
   \frac{\left(1 - \frac{2 \alpha}{3\pi} \ln \frac{m_\mu}{m_e}
   \right)^{9/2}}{1 - \frac{3 \alpha}{\pi} \ln \frac{m_\mu}{m_e}}
   = 1.00055
\ee
which multiplies $R_{e/\mu}$. This implies that the $O(\alpha^2)$
correction to $R_{e/\mu}$ is of the order of $0.05 \%$.

If we multiply our result with this
enhancement factor (see \eqn{eqn3125}), we get
\be
 \mbox{$ \D  R_{e/\mu} = (1.2356 \pm 0.0001) \cdot 10^{-4}$}
\ee
or
\be
  \mbox{$ \D \delta R_{e/\mu} = \left( - 3.74 \pm 0.01 \right)
  \% $}
\ee
 which agrees with the Marciano-Sirlin result within their error bars.
%
%
\section{Summary and Conclusions}
%
We have calculated the radiative corrections to the decays
$\tau\to M \nu_\tau$ and to $M\to l\nu_l$ ($M = \pi$ or $K$, $l = e$
or $\mu$).
The central issue of this paper was the treatment of the strong
interaction.
The amplitude of the radiative decays with the emission of a real
photon can be divided into the amplitude
for internal bremsstrahlung (IB) and the amplitude for structure
dependent radiation (SD).
In the virtual corrections one has to integrate over the momentum $k$ of
the virtual photon, and therefore all the three energy regimes of the
strong interaction have to be taken into account. For small $k^2$
we have contributions from the
long distance correction, which consists of the point meson
contribution and of the hadronic structure dependent part.
The short distance corrections, which we have calculated using the
parton model, contribute in the large $k^2$ region.

Our final result for the radiative
correction $\delta R_{\tau/\pi}$ to the ratio
$\Gamma(\tau\to\pi\nu_\tau(\gamma))/ \Gamma(\pi\to\mu\nu_\mu(\gamma))$
is
$$
  \delta R_{\tau/\pi} = \left(0.16_{\T - 0.14}^{\T+
  0.09}\right)  \%
$$ and
for the ratio
$\Gamma(\tau\to K\nu_\tau(\gamma))/ \Gamma(K\to\mu\nu_\mu(\gamma))$ we
obtain
$$
   \delta R_{\tau/K } = \left(0.90_{\T - 0.26}^{\T + 0.17}\right)
   \%
$$
Note that these numbers are calculated by summing up virtual, soft and
hard photonic corrections.

We can translate the radiative corrections into predictions for the
branching ratios
\begin{eqnarray}
   \br(\tau\to\pi\nu_\tau(\gamma) & = & (11.10 \pm 0.02) \%
   \times \left(\frac{\tau_\tau}{295.7 \unit{fs}}\right)
\nonumber \\
\nonumber \\
   \br(\tau\to K \nu_\tau(\gamma) & = & (0.737 \pm 0.005) \%
   \times \left(\frac{\tau_\tau}{295.7 \unit{fs}}\right)
\nonumber \\
\nonumber \\
   \br(\tau\to h \nu_\tau(\gamma)   & = & (11.84 \pm 0.02) \%
   \times \left(\frac{\tau_\tau}{295.7 \unit{fs}}\right)
\end{eqnarray}
where $h$ denotes the inclusive sum of pions and kaons.

For the ratio $R_{e/\mu}$ of the electronic and muonic decay modes of
the pion, we obtain a radiative correction of
\be
  \mbox{$ \D \delta R_{e/\mu} = \left( - 3.74 \pm 0.01 \right)
  \% $}
\ee
resulting in
\be
 \mbox{$ \D  R_{e/\mu} = (1.2356 \pm 0.0001) \cdot 10^{-4}$}
\ee
%
\section*{Acknowledgement}
%
One of us (M.F.) would like to thank J.H. K\"uhn for helpfull
discussions.
%
\section*{Appendix}
\label{appa}
%
Except for the tau mass we use the standard particle data of
\cite{RPP92} for masses and widths. In the case of the tau mass, we
use
\be
   m_\tau = (1777.1 \pm 0.5 ) \unit{MeV}
\ee
In this paper we use the notation of
Bjorken and Drell \cite{Bjo64}, especially
\be
   \epsilon_{0123} = + 1 \qquad
   \gamma^5 = i \gamma^0 \gamma^1 \gamma^2 \gamma^3
\ee
and the $e$ is the charge of the electron, $e < 0$.
The standard loop integrals are defined by
\begin{eqnarray}
\lefteqn{   \frac{i}{16 \pi^2} A_0 (m_0) }
\nonumber \\
  & \qquad \qquad  = & \mu^{4-D}
   \int \frac{d^D k}{(2 \pi)^D} \frac{1}{k^2 - m_0^2}
\nonumber \\
\lefteqn{   \frac{i}{16 \pi^2} [B_0 \,|\, B_\mu \,|\, B_{\mu\nu}]
   (p_1,m_0,m_1)} \nonumber \\
 & \qquad \qquad = & \mu^{4-D}
   \int \frac{d^D k}{(2 \pi)^D} \frac{[1\,|\,k_\mu\,|\,k_\mu k_\nu]}
   {(k^2 - m_0^2)[(k+p_1)^2 - m_1^2]}
\nonumber \\
\lefteqn{   \frac{i}{16 \pi^2} [C_0 \,|\,C_\mu\,|\,C_{\mu\nu}]
   (p_1,p_2,m_0,m_1,m_2)}
   \nonumber \\ & \qquad \qquad = &
 \mu^{4-D}
   \int \frac{d^D k}{(2 \pi)^D} \frac{[1\,|\,k_\mu\,|\,k_\mu k_\nu]}
   {(k^2 - m_0^2)[(k+p_1)^2 - m_1^2][(k+p_2)^2 - m_2^2]}
\nonumber \\
\nonumber \\
\end{eqnarray}
The scalar functions depend
on invariant combinations of the momenta only,
\begin{eqnarray}
   B_0(p_1,m_0,m_1) & \equiv & B_0(p_1^2,m_0,m_1)
\nonumber\\[0.5ex]
   C_0(p_1,p_2,m_0,m_1,m_2) & \equiv &
   C_0(p_1^2,(p_1-p_2)^2,p_2^2,
       m_0,m_1,m_2)
\end{eqnarray}
and the vector and tensor integrals are decomposed covariantly in
the form
\begin{eqnarray}
   B^\mu(p_1,m_0,m_1) & = & B_1 p_1^\mu
\nonumber \\[0.5ex]
   B^{\mu\nu} (p_1,m_0,m_1) & = & B_{00} g^{\mu\nu}
   + B_{11} p_1^\mu p_1^\nu
\nonumber \\[0.5ex]
   C^\mu(p_1,p_2,m_0,m_1,m_2) & = &
   C_1 p_1^\mu + C_2 p_2^\mu
\nonumber \\[0.5ex]
   C^{\mu\nu}(p_1,p_2,m_0,m_1,m_2) & = &
   C_{00} g^{\mu\nu} + C_{11} p_1^\mu p_1^\nu
   + C_{22} p_2^\mu p_2^\nu + C_{12} \left(p_1^\mu p_2^\nu
      + p_2^\mu p_1^\nu \right)
\end{eqnarray}
where
\begin{eqnarray}
   B_i  & =& B_i(p_1^2,m_0,m_1)
\nonumber \\[0.5ex]
   B_{ij} &=& B_{ij}(p_1^2,m_0,m_1)
\nonumber \\[0.5ex]
   C_i &=& C_i(p_1^2,(p_1-p_2)^2,p_2^2,
       m_0,m_1,m_2)
\nonumber \\[0.5ex]
   C_{ij} &=& C_{ij} (p_1^2,(p_1-p_2)^2,p_2^2,
       m_0,m_1,m_2)
\end{eqnarray}
%

%
\newpage
\renewcommand{\baselinestretch}{1.0}
\begin{figure}
   \caption{The three different classes of radiative corrections}
   \label{figc1}
\begin{center}
\fbox{\str \ \dots \ }
  \end{center}
\end{figure}
\begin{figure}
\caption{The function $r_{\tau/\pi}(u)$ for $\mu_{cut} = 1 \unit{GeV}$
(dashed), $\mu_{cut} = 1.5 \unit{GeV}$ (dotted)  and $\mu_{cut} =  2
\unit{GeV}$ (solid)}
\label{figwave1}
\begin{center}
\fbox{\str \ \dots \ }
  \end{center}
\end{figure}
\begin{figure}
\caption{The function $r_{\tau/K}(u)$ for $\mu_{cut} = 1 \unit{GeV}$
(dashed), $\mu_{cut} = 1.5 \unit{GeV}$ (dotted)  and $\mu_{cut} =  2
\unit{GeV}$ (solid)}
\label{figwave2}
\begin{center}
\fbox{\str \ \dots \ }
  \end{center}
\end{figure}
\begin{figure}
\caption{Radiative correction to $R_{\tau/\pi}$: Complete prediction
using the
standard parameter set (solid) and prediction for short distance plus
point pion only, ie.\ $(\delta R_{\tau/\pi})_{CPM} + (\delta
R_{\tau/\pi})_{sd}$ (dashed)}
\label{figm1}
\begin{center}
\fbox{\str \ \dots \ }
  \end{center}
\end{figure}
\clearpage
\begin{figure}
\caption{The total radiative correction $\delta R_{\tau/\pi}$
for different choices for the
parameters of the structure dependent correction: Standard choice (I)
(solid), choice (II) (dashed) and choice (III) (dotted)}
\label{figtaupi}
\begin{center}
\fbox{\str \ \dots \ }
  \end{center}
\end{figure}
\begin{figure}
\caption{The total radiative correction $R_{\tau/K}$
for different choices for the
parameters in the structure dependent correction: Standard choice (I)
(solid), choice (II) (dashed) and choice (III) (dotted)}
\label{figtauka}
\begin{center}
\fbox{\str \ \dots \ }
  \end{center}
\end{figure}
\end{document}